\documentclass[twocolumn]{aastex631}

\usepackage[flushleft]{threeparttable}
\usepackage{multirow}
\usepackage[normalem]{ulem}

\shorttitle{HST Transmission spectrum of KELT-20 b}
\shortauthors{Chachan et al.}

\begin{document}

\title{Strong NUV Refractory Absorption and Dissociated Water in the \emph{Hubble} Transmission Spectrum of the Ultra Hot Jupiter KELT-20 b}

\correspondingauthor{Yayaati Chachan}
\email{ychachan@ucsc.edu}

\author[0000-0003-1728-8269]{Yayaati Chachan}
\affiliation{Department of Astronomy and Astrophysics, University of California, Santa Cruz, CA 95064, USA}

\author[0000-0003-3667-8633]{Joshua Lothringer}
\affiliation{Space Telescope Science Institute, Baltimore, MD 21218, USA}

\author[0000-0001-9164-7966]{Julie Inglis}
\affiliation{Division of Geological and Planetary Sciences, California Institute of Technology, Pasadena, CA, 91125, USA}

\author[0000-0002-6980-052X]{Hayley Beltz}
\affiliation{Department of Astronomy, University of Maryland, College Park, MD 20742, USA}

\author[0000-0002-5375-4725]{Heather A. Knutson}
\affiliation{Division of Geological and Planetary Sciences, California Institute of Technology, Pasadena, CA, 91125, USA}

\author[0000-0002-5547-3775]{Jessica Spake}
\affiliation{Division of Geological and Planetary Sciences, California Institute of Technology, Pasadena, CA, 91125, USA}

\author[0000-0001-5578-1498]{Bjorn Benneke}
\affiliation{Department of Physics and Trottier Institute for Research on Exoplanets, Universit\'e de Montr\'eal, Montr\'eal, QC, Canada}

\author[0000-0001-9665-8429]{Ian~Wong}
\affiliation{Space Telescope Science Institute, 3700 San Martin Drive, Baltimore, MD 21218, USA}

\author[0000-0003-4408-0463]{Zafar Rustamkulov}
\affiliation{Department of Earth and Planetary Sciences, Johns Hopkins University, Baltimore, MD, USA}

\author[0000-0001-6050-7645]{David Sing}
\affiliation{Department of Earth and Planetary Sciences, Johns Hopkins University, Baltimore, MD, USA}
\affiliation{Department of Physics and Astronomy, Johns Hopkins University, Baltimore, MD USA}

\author[0000-0002-9030-0132]{Katherine A. Bennett}
\affiliation{Department of Earth and Planetary Sciences, Johns Hopkins University, Baltimore, MD, USA}

\begin{abstract}
Ultra hot Jupiters (UHJs) present a promising pathway for drawing a link between a planet’s composition and formation history. They retain both refractory and volatiles species in gas phase in their atmospheres, which allows us to place unique constraints on their building blocks. Here, we present the 0.2 - 1.7 $\mu$m transmission spectrum of KELT-20 b/MASCARA-2 b taken with the Hubble Space Telescope (HST). Unlike other UHJs around early-type stars, KELT-20 b’s orbit is well aligned with its host star’s spin axis and we test whether its distinct dynamical configuration is reflected in its composition. We observe a tremendous rise ($> 10$ scale heights) in the planet's transit depth at the near-UV wavelengths, akin to that observed for WASP-178 b and WASP-121 b, and a muted water absorption feature in the near-IR. Our retrievals indicate that the large NUV depth is driven by Fe II and/or SiO and that the water is mostly thermally dissociated. Assuming equilibrium chemistry, we obtain constraints on Z/H and O/H that indicate accretion of volatile-rich solids and/or gas. Both our low resolution spectrum and the refractory elemental ratios from \cite{Gandhi2023} suggest that nightside condensation and rainout are limited to only the most refractory species in the planet’s atmosphere. Within the precision limits of the HST spectra, no strong evidence for limb asymmetry is detected. We contextualize this lack of asymmetry by comparing to predictions from general circulation models with and without the effects of kinematic magnetohydrodynamics. Lastly, we find no major differences in the HST transmission spectra of KELT-20 b, WASP-178, and WASP-121 b despite their different dynamical configurations.

\end{abstract}

\section{Introduction} \label{sec:intro}

Relating planet formation to atmospheric composition has been a long standing goal of the astrophysics community. The proportions in which a gas-rich planet accretes gas and dust determine its final atmospheric composition. Since the composition of dust and gas varies with location in a protoplanetary disk, we expect a planet’s atmospheric composition to encode information about where it formed \citep{Oberg2011}. Ultra hot Jupiters have emerged as particularly exciting targets for relating formation to composition \citep{Lothringer2021, Chachan2023}: they are hot enough to retain both volatile and refractory elements in gas phase \citep{Kitzmann2018, Parmentier2018, Lothringer2018} and provide us with the most complete inventory possible of planetary envelope composition. Refractory elements such as iron and silicon are particularly valuable because they are invariably present in the solid phase throughout most of the protoplanetary disk. By measuring their abundance in a planet's atmosphere, we can determine the relative amount of solid-to-gas accretion during its formation \citep{Chachan2023}. When combined with volatile elements such as carbon and oxygen, ultra hot Jupiter atmospheres can shed light on the ice content of the accreted solids, the extent of their thermal and chemical processing \citep{Lichtenberg2019, Lichtenberg2021}, and as a result their formation conditions and location. The combination of refractories and volatiles enables us to lift degeneracies in these inferences that are present when only volatile element abundances can be characterized \citep{Mordasini2016, Schneider2021, Turrini2021, Molliere2022, Pacetti2022}. 

Dynamical properties of planets also provide an excellent probe of their origin \citep[see][for a review]{Dawson2018}. The spin-orbit angle (between the planet's orbital axis and the stellar spin axis) of close-in giants gives us an insight into their dynamical history. Around hot stars (above the Kraft brake $T_{\rm eff} \sim 6100$~K), the primordial spin-orbit angle is likely preserved as it is not easily damped away by dissipation inside the star \citep[e.g.,][]{Dawson2014, Zanazzi2024}. Most giant planets found around hot stars exhibit a relatively large spin-orbit misalignment and they are often on polar orbits \citep{Schlaufman2010, Winn2010}. This likely reflects a common migration pathway for these planets, potentially via dynamical interactions with a third body in the system after the dissipation of the protoplanetary disk \citep[e.g.,][]{Rasio1996, Wu2003, Wu2011, Naoz2011}. Although theory predicts that migration of a giant planet through a protoplanetary disk should result in higher solid accretion rate and final atmospheric metallicity \citep{Shibata2019, Shibata2020} compared to migration after the disk's dissipation (which would also better preserve the compositional signature of formation at a particular location), whether these predictions are true and the extent of migration's effect on composition remain largely unverified by observations.

Here, we add KELT-20 b to the small sample of ultra hot Jupiters that have been studied with HST and/or JWST. KELT-20 b orbits an exceptionally hot and bright star ($T_{\rm eff} \sim 8700$ K, $V$ mag = 7.6) every 3.5 days, and has an equilibrium temperature $T_{\rm} = 2260$ K \citep[][also known as MASCARA-2 b]{Lund2017, Talens2018}. Even more interestingly, its projected obliquity is consistent with zero, i.e., the projected stellar spin axis and the planet's orbital axis are aligned \citep[see also][]{Singh2024}. Although the 3D spin-orbit angle is currently unknown, the rapid rotational velocity of the star ($v {\rm sin}i = 115.9 \pm 3.4$ km s$^{-1}$, $\Omega / \Omega_{\rm c} = 0.4$) suggests that we are viewing KELT-20 nearly equator on \citep{Lund2017}. KELT-20 b is unique among close-in planets around hot stars as the rest of them exhibit significant spin-orbit misalignment \citep[e.g.,][]{Winn2010, Anderson2018}. This special property of KELT-20 b could therefore indicate that it formed and migrated to its current location differently than other hot Jupiters around hot stars. In particular, it may have arrived at its current location via disk migration \citep{Dawson2018}, which may be reflected in its atmospheric composition.

KELT-20 b has been extensively observed from ground based facilities as well as space telescopes. The dayside emission spectrum of KELT-20 b measured with the \emph{Hubble Space Telescope (HST)} and the \emph{Spitzer Space Telescope} show water and CO spectral features in emission and indicate an inverted temperature structure \citep{Fu2022}. This study finds that CO and water abundances seem roughly equal but notably enriched relative to solar values ($\sim 10 \times$ solar), suggesting substantial volatile accretion, and a modest C/O = $0.45 \pm 0.21$. This value is compatible with the C/O = $0.1^{+0.4}_{-0.1}$ obtained from ground-based high-resolution NIR observations \citep{Finnerty2025}. There is overwhelming evidence from ground-based high-resolution spectroscopy that refractory elements are present in gaseous phase in KELT-20 b's atmosphere \citep[Ca, Cr, Fe, Mg, Mn, Ni, Si, Ti, V;][]{Casasayas-Barris2018, Casasayas-Barris2019, Stangret2020, Nugroho2020, Hoeijmakers2020, Borsa2022, Cont2022, Bello-Arufe2022, Yan2022, Kasper2023, Gandhi2023, Stangret2024, Petz2024}. These observations imply that despite being tidally locked and having a colder night side, the refractory species in KELT-20 b's atmosphere are not entirely condensed out and cold-trapped deeper in the atmosphere \citep{Parmentier2013, Lothringer2018}. \cite{Gandhi2023} homogeneously analyzed and fit these ground-based spectra to obtain elemental ratios of several refractory elements as well as a constraint on the Fe I/H = $-0.36^{+0.35}_{-0.27}$ of the planet's atmosphere, which is significantly lower than the enrichment of C and O found by \cite{Fu2022}. 

In this work, we present \emph{Hubble Space Telescope} observations of the ultra hot Jupiter KELT-20 b in the WFC3 UVIS, G102, and G141 bandpasses (summarized in \S~\ref{sec:observations}). We obtain a comprehensive transmission spectrum of this unique planet and characterize both its refractory and volatile content. Observations with WFC3 UVIS (relatively new to exoplanet observations, \citealt{Wakeford2020, Lothringer2022, Bennett2025, Boehm2025}) allow us to probe the presence of iron and silicon oxide in KELT-20 b's atmosphere in the near-UV wavelength range, where these species have strong absorption features \citep{Lothringer2020b}. The WFC3 IR observations in the G102 and G141 bandpasses probe molecular absorption features from water. In \S~\ref{sec:extraction}, we describe our data reduction process and introduce a novel technique to correct for higher order overlap at redder wavelengths in the UVIS bandpass. Our light curve fitting procedure is presented in \S~\ref{sec:fitting}, followed by a discussion of the white light curve and the transmission spectrum in \S~\ref{sec:results}. In \S~\ref{sec:retrievals}, we fit the transmission spectrum of KELT-20 b using two different retrieval tools, \texttt{PETRA} \citep{Lothringer2020_petra} and \texttt{petitRADTRANS} \citep{molliere_2019}, and put constraints on its Z/H and O/H. In \S~\ref{sec:discussion}, we use General Circulation Models (GCMs) to constrain the strength of KELT-20 b's magnetic field, compare its transmission spectrum to those of other UHJs, derive a condensation temperature for its night side, and explore the implications of our measured abundances for its formation and migration history.

\section{Observations} \label{sec:observations}
We observed four transits of KELT-20 b with the Hubble Space Telescope using the WFC3 instrument as part of the GO 17082 program (PI: Chachan). Two of these visits were observed in the UVIS bandpass on Nov 23 2022 and Dec 31, 2022. Following the observational setup from GO 16086 \citep{Lothringer2022}, we used a subarray of size $2250 \times 590$ to minimize data volume and center the subarray readout on the zeroth order of G280 spectrum using CENTERAXIS1 = 2136 and CENTERAXIS2 = 1216. The target acquisition image was taken with the narrower F469N filter to avoid saturation. Due to an error in Phase II setup (the POSTARG Y -50" offset that is necessary for centering the target on the subarray on chip 2 was specified for the acquisition image but not the science images), the star was not centered on the sub-array during the first visit and these data are therefore unusable. The second visit was executed correctly and consisted of five HST orbits. We chose an exposure time of 23.5~s such that both the +1 and -1 orders could be used for measurements and to improve duty cycle. Although this led to partial saturation in the +1 order, we demonstrate that independent measurements of transit depths from both the +1 and -1 orders are consistent with each other. 

\begin{figure*}
    \centering
    \includegraphics[width=\linewidth]{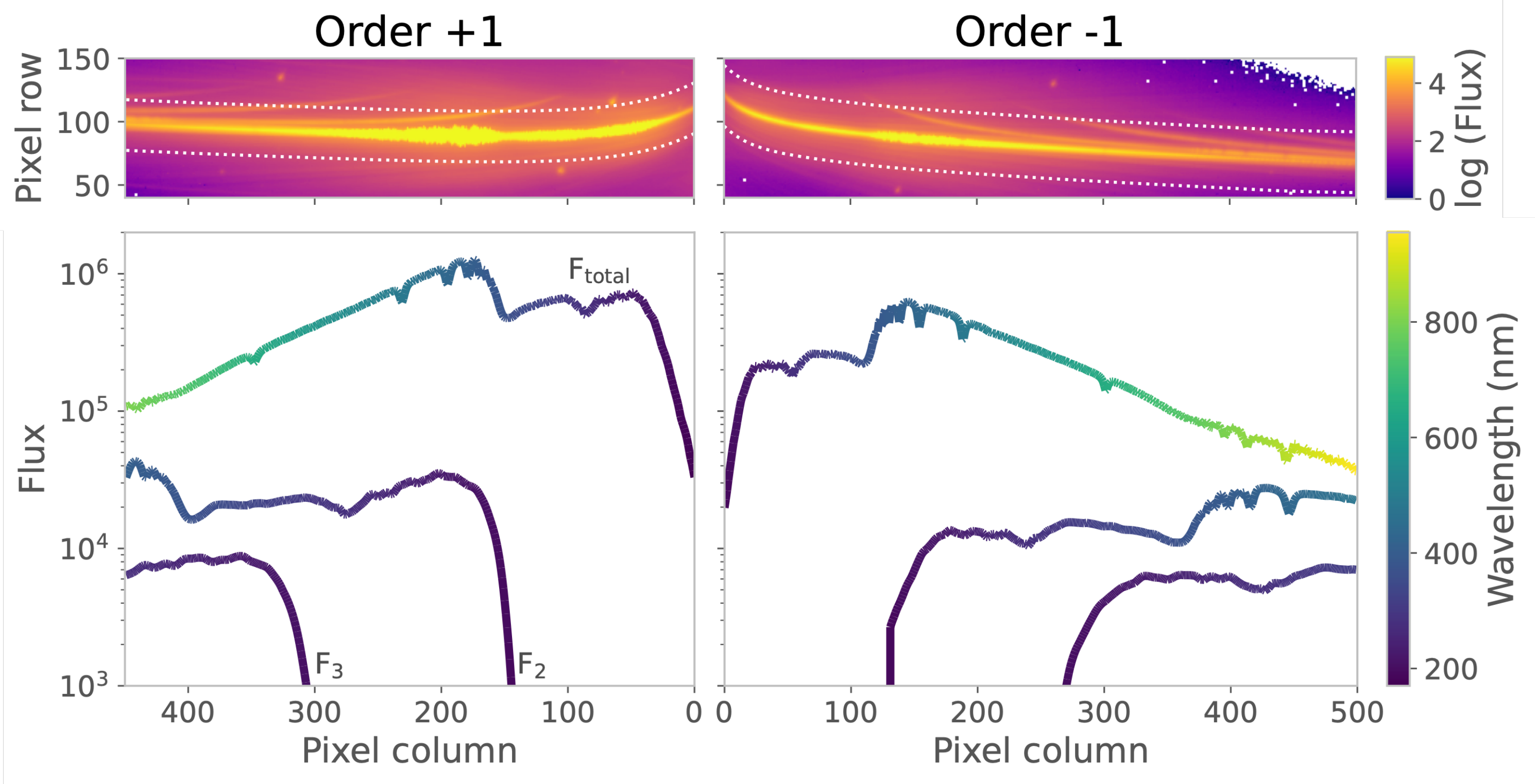}
    \caption{Top panel: Logarithm of the median flux in the traces of the two orders with the UVIS instrument after background subtraction and outlier correction for each pixel in the 3D data cube. The dotted lines mark the apertures ($\pm 20$ pixels and $\pm 24$ pixels for +1 and -1 order respectively) within which the flux is summed to obtain the 1D spectra. Higher order traces enter the aperture and contaminate the flux from the primary order.
    Bottom panel: The total flux $F_{\rm total}$ within the aperture as well as the flux in orders 2 and 3 ($F_2 = S_2 \, F_{*, \lambda_2} \Delta \lambda_2 t_{\rm exp}$ and $F_3 = S_3 \, F_{*, \lambda_3} \Delta \lambda_3 t_{\rm exp}$, see Equation~\ref{eq:intrinsic_flux}) for both the +1 and -1 orders from the UVIS data. The color of the lines indicates the wavelength of the different traces at any given pixel column. The flux in higher orders is derived by multiplying the respective calibrated sensitivities with the iteratively determined intrinsic flux of the star. At redder wavelengths for the $\pm 1$ orders, the higher orders constitute a significant fraction of the total flux.
    }
    \label{fig:trace_flux}
\end{figure*}

The third visit observed KELT-20 in the G102 bandpass on Feb 14, 2023 while the fourth visit observed it in the G141 bandpass on Feb 28, 2023. For the G102 and G141 visits, we followed the setup from prior observations of KELT-20 b's secondary eclipse in the HST program GO 16307 \citep{Fu2022}. We used scanning mode with the SQ512SUB sub-array for our observations to spread out the trace in the cross-dispersion direction, minimize read-out time, and improve duty cycle. Direct images of the star were acquired at the start of the observations with an exposure time of 0.85~s using the F126 and F164N filter for the G102 and G141 observations, respectively. The observations utilized both forward and backward scans with a scan rate of 0.702~arcsec/s. Each exposure lasted 69.6~s. Six HST orbits per visit were needed to observe the transit of KELT-20 b in the infrared bandpasses as the first orbit exhibits worse systematics and often needs to be discarded in the final analysis.

\section{Photometric \& Spectral Extraction} \label{sec:extraction}

\subsection{HST WFC3 UVIS}

We start the extraction process with the \texttt{flt} files outputted by the \texttt{calwfc3} pipeline. The reduction and extraction procedure follows the study that pioneered the use of WFC3 UVIS mode for exoplanet time series observations \citep{Wakeford2020}, with some important changes to address issues arising from the brightness of KELT-20 and our wish to utilize the -1 order trace for scientific analysis.

\textit{Background subtraction}: We use the median of each frame to estimate the background and subtract it before any further processing. 
We found that this performed comparably to estimating the background using the median and sigma clipping with \texttt{astropy} \citep{2013A&A...558A..33A, 2018AJ....156..123A} and was superior to the mode \citep{Wakeford2020} as measured by the scatter in the white light curve.

\textit{Trace \& wavelength solution}: We use the updated calibration files from \cite{Pirzkal2020} to calculate the trace location and the wavelength solution. We first determine the position of the sub-array within the full detector array using \texttt{wfc3tools}. Then, using this position as well as the position of the star in the direct image, we calculate the trace profile and wavelength solution for $\pm 1$, $\pm 2$ and $\pm 3$ orders. The calibrated trace and wavelength profiles are most accurate for the first exposure after the direct image. Due to slight shifts in the trace position over subsequent exposures, we find that using an empirically fitted $y$-trace (eighth order polynomial function of the $x$-trace) for spectral extraction yields a significant reduction in the residuals in the white light curve and we therefore use these empirical trace profiles in our final analysis.

\textit{x and y position shifts}: There are sub-pixel shifts in the position of the star and the trace over the course of the visit. We calculate these shifts by fitting the positions of four different stars on the detector using a 2D Gaussian with \texttt{photutils} \citep{Bradley2018}. We find that the position estimates agree well with each other so we adopt the position of the brightest star (positioned at [1855, 138] in the subarray) as our marker for the shifts in the $x$ and $y$ direction. However, to test whether our position movements derived from field stars matches the spectrum's movement, we devise a method to calculate the shifts using the H$\beta$ and H$\gamma$ lines. The $x$ shifts are difficult to estimate directly from the stellar spectrum because KELT-20 is an A star and its spectrum is dominated by broad Balmer lines of hydrogen. As a result, cross-correlation with a \texttt{PHOENIX} spectrum does not yield an accurate estimate of the trace's movement.  We locally model these lines as a Gaussian plus a linear trend and the location of the Gaussian's center is used to calculate the shifts in $x$ position. These estimates agree very well with our estimate from field stars' positions. For each exposure, we shift the spectrum according to the calculated $x$ position by linearly interpolating the spectrum and calculating the flux values on a standard pixel grid.

\textit{Outlier correction}: We use a 2-step outlier correction method. In the first step, we correct for outliers in time for each pixel by building a time-median image of the detector sub-array, finding pixels that deviate $5.5 \sigma$ and $4.5 \sigma$ (for order +1 and -1 respectively) away from the median, and replacing them with the median of the pixel time series. These $\sigma$ thresholds yielded ideal outlier correction in the 2D spectral time series. A higher $\sigma$ threshold is chosen for order +1 to prevent aggressive outlier correction. Median fluxes for the two orders after this step are shown in Figure~\ref{fig:trace_flux}. We do not utilize any spatial outlier correction methods, as the bright and extended trace in the +1 order renders this approach ineffective for our data. We remove the remaining outliers by working with the 1D spectral time series instead. For each 1D spectrum, we calculate the median and standard deviation from the surrounding six exposures and replace any pixels in the chosen spectrum that deviate by more than $6 \sigma$ with the median. These two procedures combined remove nearly all visible outliers without over-correcting or introducing artifacts.

\textit{Photometric and spectral time series}: We sum the flux within an aperture of size $\pm 20$ pixels for +1 order and $\pm 24 $ pixels for -1 order centered on the trace (aperture extent shown in Figure~\ref{fig:trace_flux}). For order +1, the aperture size was chosen to be as large as possible while excluding contaminating light from a nearby star. We choose a slightly larger aperture for the dimmer order -1 trace to include more photons and reduce the residuals in the white light curve. Accounting for the $y$ position shifts in the trace improves the estimate of the star's flux and reduces the scatter in the white light curve. We add contributions from partial pixels in the $y$ direction by multiplying the fraction of a pixel that lies within the aperture with the pixel's flux count. For the spectroscopic light curves, we calculate the pixel extents corresponding to wavelength bins separated by 20 nm between 200 nm and 800 nm and sum the 1D spectra between these pixel values. Contributions from partial $x$ pixels are included using the same procedure outlined above for partial $y$ pixels.

\textit{Contamination from higher orders}: The UVIS spectral traces overlap at redder wavelengths and this introduces contamination that biases the measured transit depths (Figure~\ref{fig:trace_flux}). Due to partial spectral saturation and position shifts that significantly change the flux of the star, we are unable to construct a reliable spatial profile of the trace in the cross-dispersion direction and use optimal extraction to separate the orders. However, the effect of contamination is not as severe as that from a stray star because the planet transit also dims the higher orders. The overlapping order's impact on the transit depth in given wavelength range due to order n can be estimated as follows:
\begin{equation}
    \Delta \delta_{\rm n} = (\delta_{1} - \delta_{\rm n}) \frac{F_{\rm n}}{F_{\rm total}}
\end{equation}
where $\Delta \delta_{\rm n}$ is the correction to the transit depth in order 1 due to contamination from order n, $\delta_{1}$ and $\delta_{\rm n}$ are the transit depths in the respective wavelength ranges of order 1 and n in the overlap region. $F_{\rm n}$ is the flux from the contaminating order n and $F_{\rm total}$ is the total flux within the overlap region (Figure~\ref{fig:trace_flux}).

\begin{figure}
    \centering
    \includegraphics[width=\linewidth]{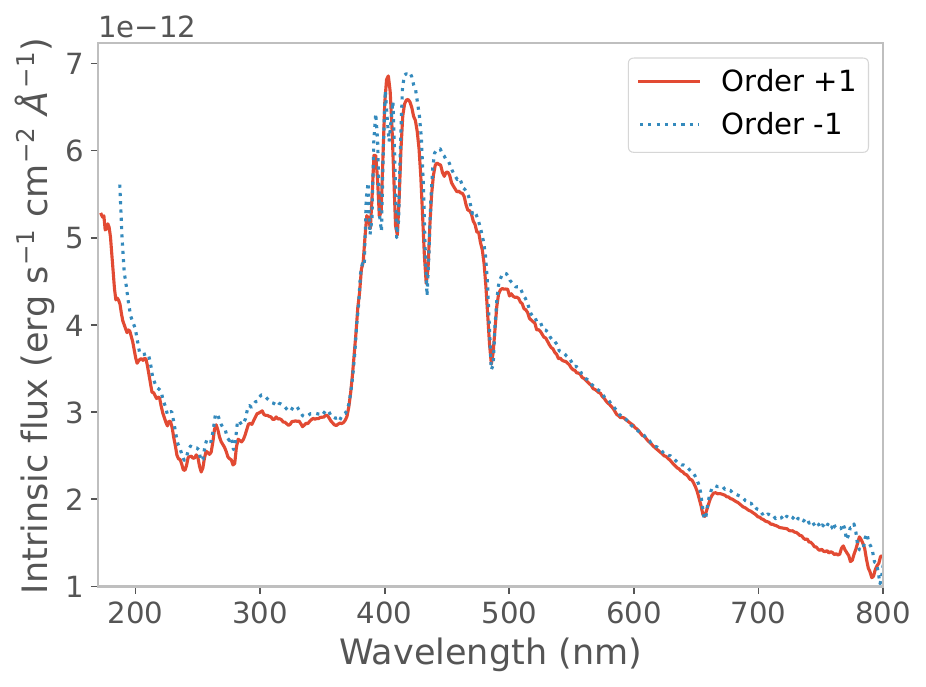}
    \caption{The iteratively determined intrinsic flux of the star KELT-20 from the order +1 and -1 traces.}
    \label{fig:intrinsic_flux}
\end{figure}

To calculate this correction, we need to measure the transit depths for the different wavelength ranges that overlap as well as the fractional contribution of the contaminating order to the total flux inside the aperture. We use sensitivity calibrations files from \cite{Pirzkal2020} to iteratively determine the underlying `intrinsic' flux $F_*$ of the observed star (Figure~\ref{fig:intrinsic_flux}). The total flux within the aperture can be written as the sum of contributions from different orders:
\begin{equation}
    \frac{F_{\rm total}}{t_{\rm exp}} = S_1 \, F_{*, \lambda_1} \Delta \lambda_1 + S_2 \, f_2 \, F_{*, \lambda_2} \Delta \lambda_2 + S_3 \, f_3 \, F_{*, \lambda_3} \Delta \lambda_3,
    \label{eq:intrinsic_flux}
\end{equation}
where $t_{\rm exp} = 23.5$~s is the exposure time, $S_{\rm n}$ is the sensitivity of order n, $f_{\rm n}$ is the fraction of flux of order n that is within the aperture (assumed to be 1 for the first order), $\Delta \lambda_{\rm n}$ is the bin width in $\AA$ for order n, and $F_{*, \lambda_{\rm n}}$ is the intrinsic flux of the star in the wavelength range corresponding to order n. This procedure allows us to determine the absolute flux contributions of the different orders. In each column, we calculate $f_{\rm n}$ assuming that the order's PSF is well represented by a Gaussian. The width of the Gaussian is calculated by locally fitting the cross-dispersion profile centered on the contaminating order with a Gaussian + a straight line for a range of columns near the intersecting region (Figure~\ref{fig:trace_flux}, e.g., columns $145 - 170$ for +2 order) and by taking the median of the calculated Gaussian profile widths. For any column in which the contaminating order's central trace is 3~$\sigma$ away from the aperture, we assume the contribution of the order's flux to be either zero (trace outside aperture) or one (trace inside aperture). 

\textit{Construction of white light curve}: We sum the flux between $173 -  804$ nm (450 columns) for the +1 order and $187 -  956$ nm (500 columns) for -1 order. These wavelength ranges are chosen by varying the pixel extent in steps of 50 columns and minimizing the residuals in the white light curve. We choose a wider wavelength extent for the lower sensitivity -1 order to include more photons in the white light curve.

\subsection{HST WFC3 G102 \& G141}

\begin{figure}
    \centering
    \includegraphics[width=\linewidth]{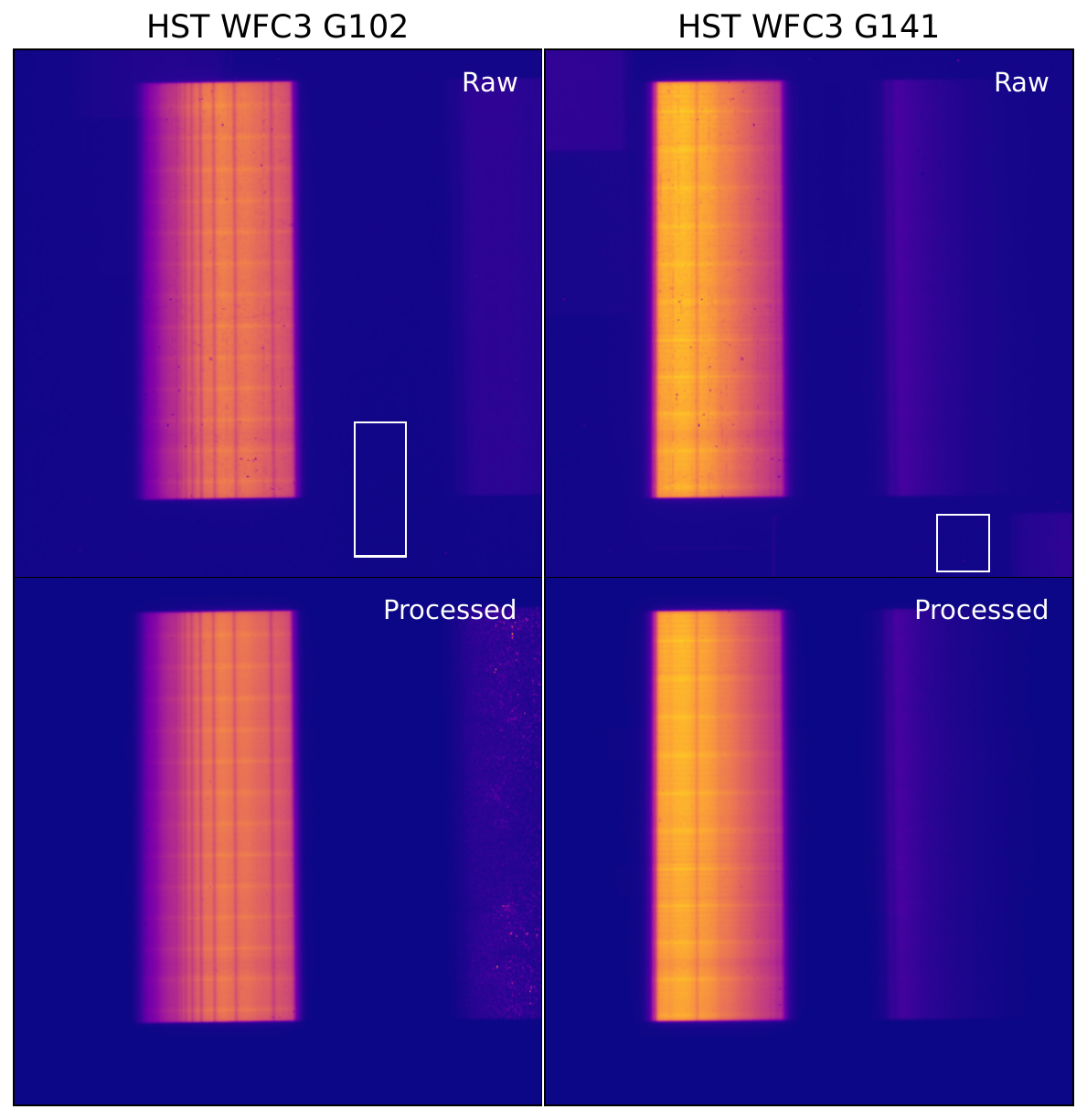}
    \caption{Raw and processed images of the first exposures of the G102 and G141 visits. The white rectangles mark the regions from which the background count is estimated.}
    \label{fig:ir_obs}
\end{figure}

We use the \texttt{ExoTEP} pipeline to reduce our observations of KELT-20 in the G102 and G141 bandpasses. This pipeline has been extensively described in prior publications \citep{Benneke2019, Chachan2019, Chachan2020, Wong2020} and we therefore restrict our discussion to key steps and choices in the data reduction process. We start with bias- and dark-corrected \texttt{ima} images obtained from the \texttt{calwfc3} pipeline (top panels, Figure~\ref{fig:ir_obs}). These images consist of 6 nondestructive reads (including the zeroth read) from the detector and we build difference images by subtracting consecutive reads. 

We choose a rectangular region that is not contaminated by spectra from neighboring stars to estimate the background level (Figure~\ref{fig:ir_obs}). For G141 this region is in x = [380, 430] and y = [5, 60] and for G102, it is in x = [330, 380] and y = [20, 150]. We remove outliers that are $> 3 \sigma$ away from the median pixel count in this region and take the median of the remaining pixels as our background estimate, which is then subtracted from each difference image.

The vertical extent of the trace along the scan direction is set by the location at which flux value falls to 20\% of the peak value. An additional buffer of 15 pixels is added to this value to capture the remaining flux. We vary this value from $10-20$ and find that the results are not appreciably sensitive to this choice. After compiling the difference images into a combined image, we sum the spectrum along the cross-dispersion direction to obtain a 1D column added spectrum for each exposure. The WFC3 instrument is very stable but small pointing variations lead to minuscule shifts in the spectra from exposure to exposure. We find these shifts along the dispersion direction relative to the first exposure using template matching.

We obtain the wavelength solution using the trace calibration coefficients given in \citet{Kuntschner2009, Kuntschner2009a} and the flat-field image from calibration files from \cite{Kuntschner2011}. We flat-field all frames using the method given in \cite{Wilkins2014}. Following flat-fielding, we correct for outliers in the spatial domain using a $11 \times 11$ kernel centered on each pixel, finding outliers above $6 \sigma$, and replacing them with the mean value of pixel flux count within the kernel. This step is repeated twice. For outlier correction in the temporal domain, we create a median and standard-deviation image of the whole stack, find pixels that deviate more than $5 \sigma$ away from the median and replace by the median.

Finally, we extract spectra from the 2D images, where we sum the flux within a certain wavelength range while accounting for the fact that the line demarcating these wavelength regimes is not vertical but has a slight slope. We sum partial pixel contributions by fitting the flux within a $3 \times 3$ kernel with a second order polynomial in $x$ and $y$.

\section{Light Curve Fitting} \label{sec:fitting}

\subsection{Individual white light curve fit}

We commence by fitting each visit's white light curve separately using \texttt{batman} \citep{Kreidberg2015}. The astrophysical parameters that are allowed to vary are the mid-transit time $T_0$, the impact parameter $b$, the normalized orbital distance of the planet $a/R_*$, and the planet-to-star radius ratio $R_{\rm p} / R_*$. The best fit values from these individual fits are used to set the initial values for all the astrophysical and systematic model parameters in the joint fit.

\begin{figure}
    \centering
    \includegraphics[width=\linewidth]{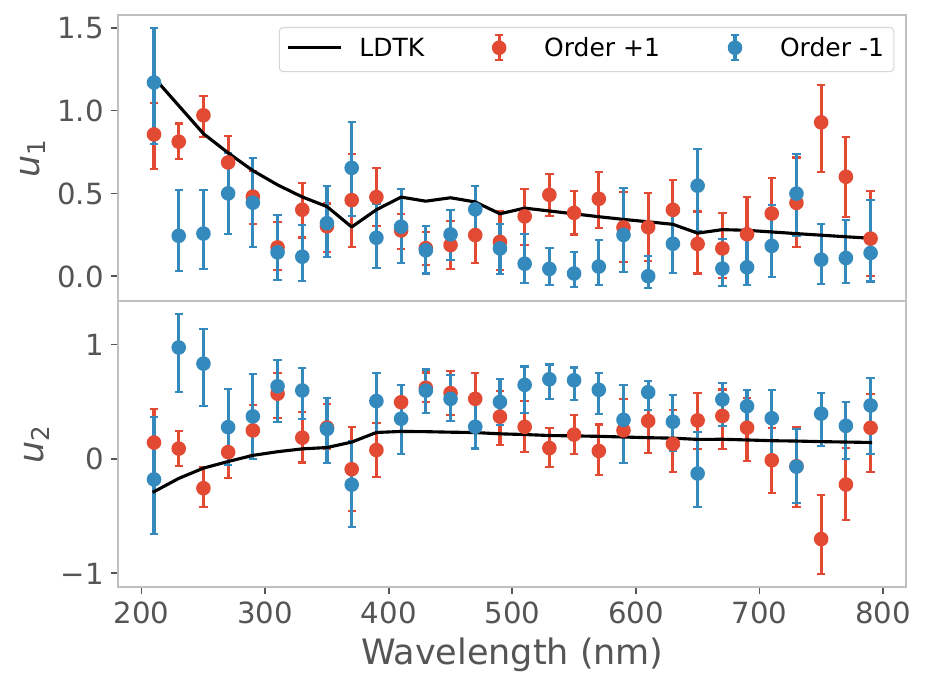}
    \caption{Fitted quadratic limb darkening coefficients for the two UVIS orders compared to model quadratic coefficients from \texttt{LDTK} \citep{Parviainen2015}.}
    \label{fig:uvis_ldc}
\end{figure}

The UVIS G280 observations are extremely stable and even the raw light curve exhibits relatively little evidence for instrumental systematics. We find that fitting for a linear function of time and the $y$-position of the trace adequately correct for the systematics effects of the telescope:
\begin{equation}
    S_{\rm WLC, G280} = c + a \, t_{\rm v} + m_y \, (y - y_0),
    \label{eq:wlc_g280}
\end{equation}
where $c$, $a$, and $m_y$ are the systematic model parameters to be fitted, $t_{\rm v}$ is the time elapsed since the beginning of the visit, and $y_0$ is the $y$-position of the first exposure. High order polynomial functions and functions of the $x$-position do not further improve the white light curve fit as measured by the Bayesian Information Criterion (BIC). 

\begin{table*}
	\begin{threeparttable}
    	\caption{Joint White Light Curve Fit Results} \label{table:wlc_fits}
    	\begin{center}
        	\begin{tabular*}{\textwidth}{@{\extracolsep{\fill}} c c c c c}
        		\hline \hline
        		Parameter & Instrument & Band pass ($\mu$m) & Prior & Value \\ \hline 
        		Transit center time $T_0$ (BJD$_{\mathrm{TDB}}$) & -- & -- &  & $2459945.412757_{-0.000037}^{+0.000037}$  \\
        		Impact parameter $b$ & -- & -- & $\mathcal{U}(0, 1)$ & $0.5132_{-0.0057}^{+0.0055}$ \\
        		Relative semi-major axis $a/R_*$ & -- & -- & -- & $7.457_{-0.033}^{+0.034}$ \\
        		Inclination\textsuperscript{a} $i$ & -- & -- &  & $ 86.054_{-0.059}^{+0.061} $ \\
                Planet radius, $R_p/R_*$ & UVIS G280 (+1 \& -1) & 0.2 - 0.8\textsuperscript{b} &  & $0.11605_{-0.00018}^{+0.00018}$     \\
                Planet radius, $R_p/R_*$ & G102 & 0.8 - 1.1 &  &  $0.114704_{-0.000072}^{+0.000070}$       \\
        		Planet radius, $R_p/R_*$ & G141 & 1.1 - 1.7 & &  $0.114688_{-0.000063}^{+0.000062}$       \\           
        		Limb darkening coefficient $u_1$ & UVIS G280 +1 & 0.17 - 0.80 & $\mathcal{U}(-0.1, 2)$ & $0.465_{-0.068}^{+0.065}$ \\
        		Limb darkening coefficient $u_2$ & UVIS G280 +1 & 0.17 - 0.80 & $\mathcal{U}(-2, 2)$ & $0.202_{-0.092}^{+0.097}$ \\
        		Limb darkening coefficient $u_1$ & UVIS G280 -1 & 0.19 - 0.96 & $\mathcal{U}(-0.1, 2)$ & $0.259_{-0.073}^{+0.072}$ \\
        		Limb darkening coefficient $u_2$ & UVIS G280 -1 & 0.19 - 0.96 & $\mathcal{U}(-2, 2)$ & $0.42_{-0.10}^{+0.10}$ \\
                Limb darkening coefficient $u_1$\textsuperscript{c} & G102 & 0.8 - 1.1 & $\mathcal{U}(-2, 2)$ & $0.1583_{-0.0065}^{+0.0064}$ \\
                Limb darkening coefficient $u_1$\textsuperscript{c} & G141 & 1.1 - 1.7 &  $\mathcal{U}(-2, 2)$ & $0.1395_{-0.0044}^{+0.0044}$ \\
        		\hline
        	\end{tabular*}
        	 \begin{tablenotes}
             \small
			 \item {\bf Notes.} 
			 \item \textsuperscript{a}{Calculated from posteriors for $b$ and $a/R_*$.}
             \item \textsuperscript{b}{The exact wavelength range is $173 -  804$ nm for +1 order and $187 -  956$ nm for -1 order. Separately fitted $R_p/R_*$ for the two orders are consistent with a single $R_p/R_*$ fitted to both of them so we adopt the latter choice. The white light curve depth is not corrected for contamination from higher UVIS orders.}
             \item \textsuperscript{c}{The limb darkening coefficient $u_2$ is obtained from \texttt{LDTK} and fixed to 0.142 and 0.110 for the G102 and G141 bandpasses respectively.}
			\end{tablenotes}
    	\end{center}
	\end{threeparttable}
\end{table*}

For the G102 and G141 observations, the first orbit and the first exposure of each orbit are discarded as they display worse systematics effects compared to the rest of the light curve. We adopt the exponential ramp model to remove the systematics of these observations: 
\begin{equation}
    S_{\rm WLC, G141, f/b} = [c_{\rm f/b} + v \; t_{\rm v}] [1 - e^{a_{\rm f/b} \, t_{\rm orb} - b - D_t}]
\end{equation}
The normalizing term $c_{\rm f/b}$ and the ramp rate $a_{\rm f/b}$ are allowed to be different for the forward and backward scans (as indicated by the subscripts) but the rest of the model parameters ($v, b, D_t$) are common to both scan directions. The term $D_t$ is a vector of length $t_{\rm v}$ and it allows us to fit orbit-specific ramp delays to the observations. For both the G102 and G141 observations, a ramp delay is added to the first orbit as it exhibits a different ramp profile than subsequent orbits. For the G102 data, an additional second ramp delay is added for the last orbit as it significantly improves the fit and is favored by BIC. 

In the UVIS bandpass, we fit for both coefficients of the quadratic limb darkening law to ensure sufficient flexibility in the modeling of the stellar intensity profile. Given the strong impact of limb darkening in the NUV and optical bandpasses, we find that the light curves are able to sufficiently constrain the limb darkening parameters. The modeled limb darkening coefficients (LDTk, \citealt{Parviainen2015}) are in reasonable agreement with the fitted values (Figure~\ref{fig:uvis_ldc}), with some differences possibly due to contamination from higher-order spectra. Fixing either just one of the coefficients or both coefficients leads to biases in the transit depths \citep[see also][]{Coulombe2024}. Fixing only one coefficient also does not significantly reduce the uncertainties on the transit depths so we fit for both coefficients, which enables us to marginalize over incomplete knowledge of the stellar limb darkening in the UVIS bandpass.  For the G102 and G141 bandpasses, we still use the quadratic limb darkening law but fit for only one parameter while fixing the other to the value obtained from \texttt{LDTK}. 
When we fit for both coefficients for the G102 and G141 bandpasses, we find that they are poorly constrained by the data and comfortably encompass the \texttt{LDTK} values. This is due to the weaker effect of limb darkening in the NIR G102 and G141 bandpasses and the periodic gaps in the light curves, scenarios in which \cite{Coulombe2024} support the use of theoretical limb darkening coefficients. Fixing one coefficient and fitting for the other allows for some flexibility in the light curve modeling without under-constraining the fit and the fitted coefficients match the LDTk model values very well. This strategy is similar to the one adopted by the \emph{JWST} ERS program for precision transit light curve modeling \citep{Rustamkulov2023, Alderson2023, Ahrer2023}.

\begin{figure*}
    \includegraphics[width=\linewidth]{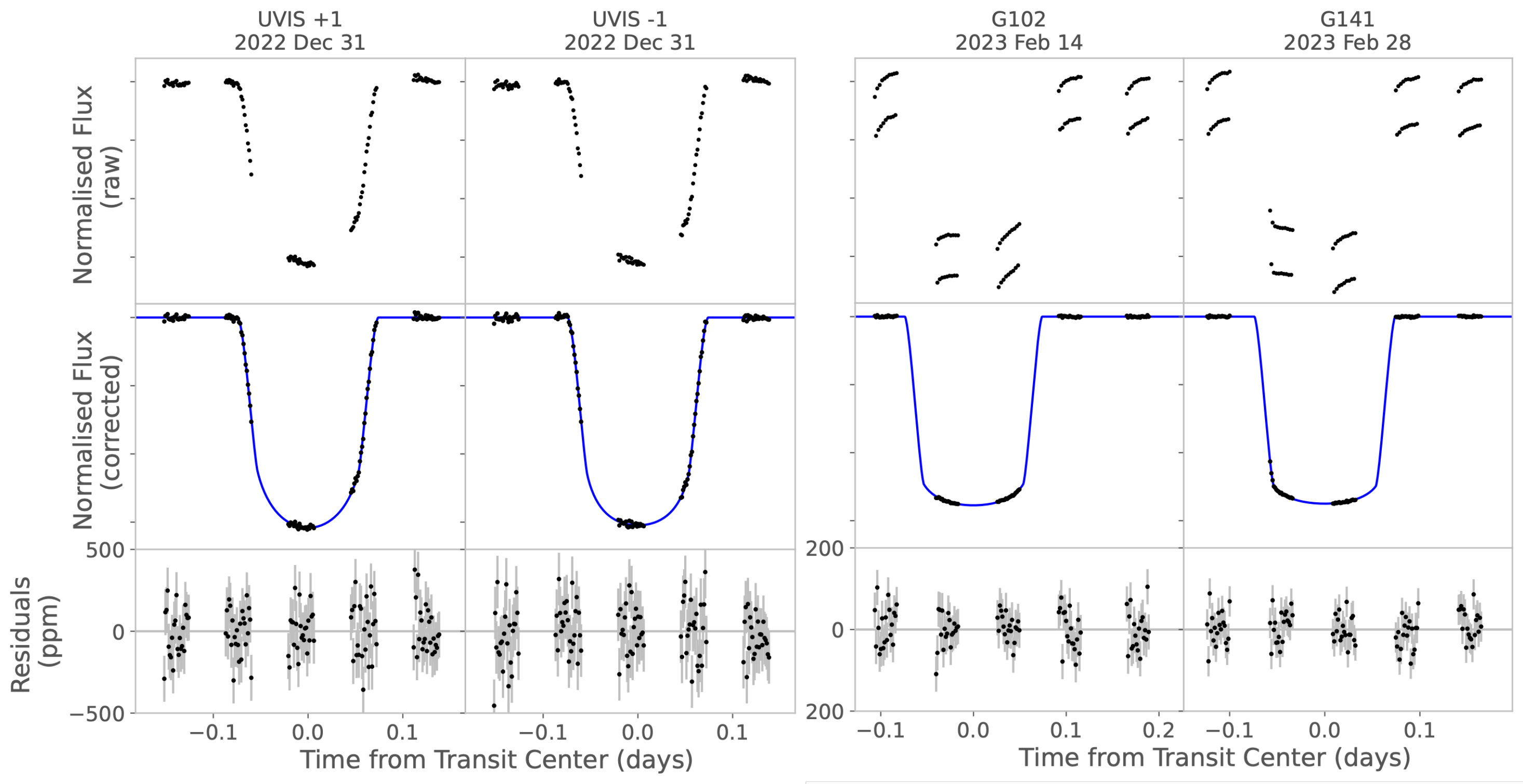}
    \caption{The figure shows the raw white light curves (top panel), transit models and white light curves with the systematics removed (middle panel), and the residuals (bottom panel). For the G102 and G141 visits, the residuals window spans -200 to 200 ppm ($2.5 \times$ Zoom). The UVIS transits are significantly deeper than the G102 and G141 transits.}
    \label{fig:wlcs}
\end{figure*}

\subsection{Joint white light curve fit} \label{sec:wlc}

We perform a joint fit of the two light curves from the +1 and -1 orders from UVIS, the G102 light curve, and the G141 light curve. This enables us to place tighter constraints on the astrophysical parameters ($T_0$, $a/R_*$, $b$) that are assumed to be the same for each visit and fit a white light curve transit depth for each bandpass. The orbital period is fixed to the value ($P = 3.47410024$ days) reported in \cite{Ivshina2022}. We fit separate quadratic limb darkening coefficients to the +1 and -1 order UVIS white light curves as they span different wavelength ranges and are contaminated to differing extents by higher order traces. Allowing for different transit depths for the two UVIS light curves yields values for $R_{\rm p} / R_*$ that are consistent with each other so we just fit a single transit depth to the two light curves. 

With the astrophysical and visit specific instrumental parameters, the joint fit has a total of 37 parameters. We use \texttt{emcee} \citep{Foreman-Mackey2012} to fit the data in a Bayesian framework with a 5,000 step burn-in. The positions at the end of the burn-in are used to initialize a 20,000 step fit and the first 1,000 steps are discarded before constructing the posterior distributions. Convergence is confirmed by examining the chain plots and calculating the effectively independent samples ($> 3000$ for all astrophysical parameters) using the integrated autocorrelation time using \texttt{emcee}.

\begin{figure}
    \centering
    \includegraphics[width=\linewidth]{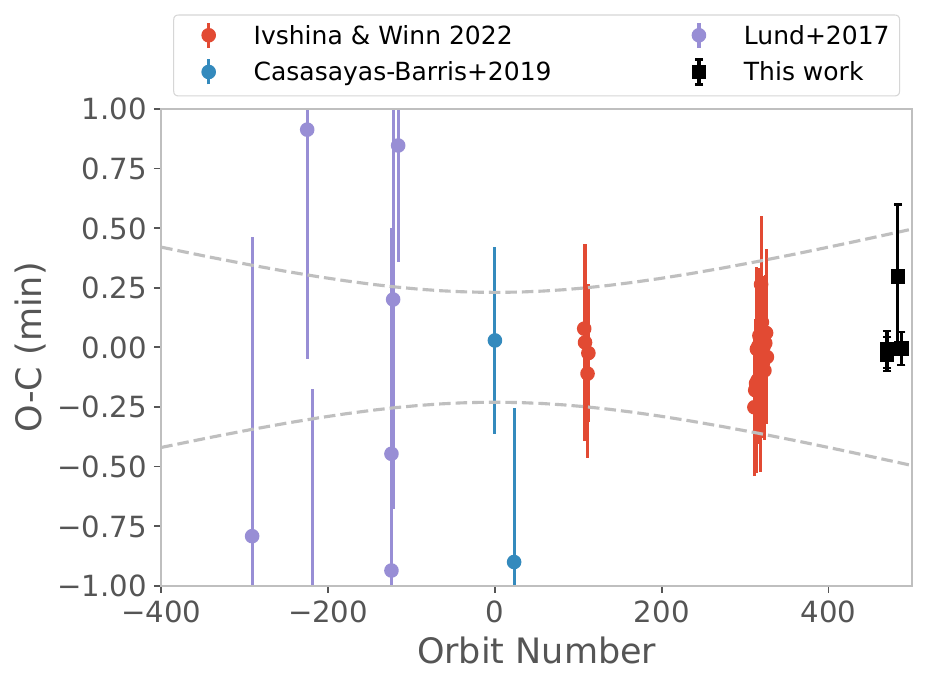}
    \caption{Observed minus calculated mid-transit times of KELT-20 b. The period and $T_0$ used to calculate the uncertainty region (grey lines) are from \cite{Ivshina2022}. Mid-transit times from this work as shown in black squares. $T_0$ of the UVIS +1 and -1 order transits as well as the G141 transit are in exquisite agreement with the predicted $T_0$. The WFC3 G102 visit does not cover either ingress or egress and its measured $T_0$ has the largest uncertainty out of the four light curves in our study.}
    \label{fig:O_C}
\end{figure}

\subsection{Spectroscopic light curve fit}
We fit the spectroscopic light curves while keeping $T_0$, $b$, and $a/R_*$ fixed to their best fit values from the joint white light curve fit. The spectroscopic bins span 20 nm for the UVIS bandpass and 30 nm for the G102 and G141 bandpasses. For all three bandpasses, we find that fitting the full systematics model to each spectroscopic light curve yields the cleanest residuals. For each light curve, we fit for $R_{\rm p} / R_*$, limb darkening coefficients, and the systematics model parameters with 10,000 steps. For G102 and G141, we use the same systematics model as the white light curve. For the G280 bandpass, we add a term linear in the $x$-position shifts to Equation~\ref{eq:wlc_g280} as it yields marginally better fits:
\begin{equation}
    S_{\rm Wave LC, G280} = c + a \, t_{\rm v} + m_y \, (y - y_0) + m_x \, (x - x_0).
    \label{eq:wavelc_g280}
\end{equation}

\begin{table}
	\begin{threeparttable}
    	\caption{HST WFC3 G102 \& G141 Spectroscopic Depths} \label{table:g102_g141_fits}
        	\begin{tabular*}{\linewidth}{l c c c c}
        		\hline \hline
        		Bandpass ($\mu$m) & $R_{\rm p} / R_*$ & $\sigma_{R_{\rm p} / R_*}$ & $\delta$ (ppm) & $\pm 1 \sigma$ (ppm) \\ \hline 
                \textbf{G102} &  &  &  & \\
                0.80-0.83 & 0.11481 & 0.00023 & 13181 & 53 \\
                0.83-0.86 & 0.11496 & 0.00020 & 13217 & 46 \\
                0.86-0.89 & 0.11491 & 0.00020 & 13204 & 46 \\
                0.89-0.92 & 0.11464 & 0.00016 & 13142 & 36 \\
                0.92-0.95 & 0.11475 & 0.00014 & 13167 & 32 \\
                0.95-0.98 & 0.11497 & 0.00013 & 13217 & 30 \\
                0.98-1.01 & 0.11472 & 0.00013 & 13160 & 30 \\
                1.01-1.04 & 0.11468 & 0.00014 & 13152 & 31 \\
                1.04-1.07 & 0.11492 & 0.00013 & 13206 & 29 \\
                1.07-1.10 & 0.11470 & 0.00013 & 13156 & 31 \\
                1.10-1.13 & 0.11475 & 0.00015 & 13166 & 34 \\
                \textbf{G141} &  &  &  & \\
                1.10-1.13 & 0.11474 & 0.00014 & 13165 & 33 \\
                1.13-1.16 & 0.11480 & 0.00016 & 13178 & 36 \\
                1.16-1.19 & 0.11476 & 0.00016 & 13169 & 36 \\
                1.19-1.22 & 0.11494 & 0.00015 & 13212 & 35 \\
                1.22-1.25 & 0.11489 & 0.00016 & 13199 & 36 \\
                1.25-1.28 & 0.11458 & 0.00015 & 13129 & 35 \\
                1.28-1.31 & 0.11424 & 0.00014 & 13051 & 32 \\
                1.31-1.34 & 0.11467 & 0.00016 & 13150 & 36 \\
                1.34-1.37 & 0.11493 & 0.00020 & 13209 & 46 \\
                1.37-1.40 & 0.11470 & 0.00017 & 13156 & 39 \\
                1.40-1.43 & 0.11511 & 0.00015 & 13251 & 35 \\
                1.43-1.46 & 0.11511 & 0.00016 & 13249 & 38 \\
                1.46-1.49 & 0.11473 & 0.00020 & 13163 & 46 \\
                1.49-1.52 & 0.11484 & 0.00020 & 13188 & 45 \\
                1.52-1.55 & 0.11449 & 0.00018 & 13108 & 41 \\
                1.55-1.58 & 0.11478 & 0.00022 & 13175 & 50 \\
                1.58-1.61 & 0.11482 & 0.00020 & 13183 & 46 \\
                1.61-1.64 & 0.11427 & 0.00020 & 13058 & 45 \\
                1.64-1.67 & 0.11521 & 0.00024 & 13273 & 55 \\
        		\hline
        	\end{tabular*}
	\end{threeparttable}
\end{table}

\begin{table*}
	\begin{threeparttable}
    	\caption{HST UVIS +1 order Spectroscopic Depths} \label{table:uvis_p1_fits}
    	\begin{center}
        	\begin{tabular*}{\textwidth}{@{\extracolsep{\fill}} c c c c c c c c c c c}
        		\hline \hline
                \multicolumn{5}{c}{+1 order} & 
                \multicolumn{3}{c}{+2 order} & 
                \multicolumn{3}{c}{+3 order}\\
                \cline{1-5} \cline{6-8} \cline{9-11}
        		Bandpass & $\delta_1$ & $\pm 1 \sigma$ & $\delta_{\rm 1, corrected}$ & $\pm 1 \sigma$ &  Bandpass & $\delta_2$\textsuperscript{a} & $\pm 1 \sigma$ & Bandpass & $\delta_3$\textsuperscript{a} & $\pm 1 \sigma$ \\ 
                (nm) & (ppm) & (ppm) & (ppm) & (ppm) & (nm) & (ppm) & (ppm) & (nm) & (ppm) & (ppm) \\ 
                \hline 
                200-220 & 14107 & 163 &  -- & -- &  -- & -- & -- &  -- & -- & --  \\
                220-240 & 14518 & 86 &  -- & -- &  -- & -- & -- &  -- & -- & --  \\
                240-260 & 14699 & 104 &  -- & -- &  -- & -- & -- &  -- & -- & --  \\
                260-280 & 14150 & 128 &  -- & -- &  -- & -- & -- &  -- & -- & --  \\
                280-300 & 13932 & 105 &  -- & -- &  -- & -- & -- &  -- & -- & --  \\
                300-320 & 13478 & 93 &  -- & -- &  -- & -- & -- &  -- & -- & --  \\
                320-340 & 13579 & 97 &  -- & -- &  -- & -- & -- &  -- & -- & --  \\
                340-360 & 13453 & 98 &  -- & -- &  -- & -- & -- &  -- & -- & --  \\
                360-380 & 13083 & 174 & 13066 & 174 & 190-199 & 13999 & 265 &  -- & -- & --  \\
                380-400 & 13204 & 108 & 13175 & 109 & 199-208 & 13967 & 235 &  -- & -- & --  \\
                400-420 & 13197 & 74 & 13155 & 75 & 208-217 & 14083 & 201 &  -- & -- & --  \\
                420-440 & 13171 & 65 & 13102 & 66 & 217-227 & 14292 & 127 &  -- & -- & --  \\
                440-460 & 13168 & 81 & 13078 & 82 & 227-236 & 14621 & 152 &  -- & -- & --  \\
                460-480 & 13304 & 105 & 13201 & 105 & 236-245 & 14867 & 136 &  -- & -- & --  \\
                480-500 & 13192 & 111 & 13092 & 112 & 245-254 & 14539 & 149 &  -- & -- & --  \\
                500-520 & 13390 & 100 & 13324 & 101 & 254-264 & 14318 & 133 &  -- & -- & --  \\
                520-540 & 13242 & 94 & 13183 & 94 & 264-273 & 14066 & 147 &  -- & -- & --  \\
                540-560 & 13316 & 91 & 13244 & 92 & 273-283 & 14346 & 159 &  -- & -- & --  \\
                560-580 & 13303 & 108 & 13243 & 109 & 283-293 & 13982 & 140 &  -- & -- & --  \\
                580-600 & 13192 & 126 & 13133 & 128 & 293-302 & 13718 & 133 &  -- & -- & --  \\
                600-620 & 13242 & 125 & 13179 & 127 & 302-312 & 13569 & 132 & 206-212 & 14204 & 227 \\
                620-640 & 13354 & 127 & 13289 & 130 & 312-322 & 13571 & 119 & 212-219 & 14053 & 182 \\
                640-660 & 13104 & 123 & 12967 & 127 & 322-332 & 13302 & 109 & 219-225 & 14393 & 193 \\
                660-680 & 13401 & 148 & 13236 & 154 & 332-342 & 13841 & 151 & 225-231 & 14323 & 191 \\
                680-700 & 13274 & 151 & 13045 & 158 & 342-352 & 13543 & 138 & 231-238 & 14878 & 151 \\
                700-720 & 13751 & 151 & 13713 & 159 & 352-363 & 13235 & 154 & 238-244 & 14829 & 168 \\
                720-740 & 13439 & 187 & 13312 & 202 & 363-373\textsuperscript{b} & 13095 & 251 & 244-250 & 14695 & 176 \\
                740-760 & 13601 & 201 & 13607 & 225 & 373-384\textsuperscript{b} & 13247 & 153 & 250-257 & 14338 & 176 \\
                760-780 & 13317 & 183 & 13129 & 220 & 384-395\textsuperscript{b} & 13314 & 130 & 257-263 & 14405 & 172 \\
                780-800 & 13603 & 231 & 13719 & 302 & 395-407\textsuperscript{b} & 13374 & 185 & 263-270 & 13759 & 166 \\
        		\hline
        	\end{tabular*}
            \begin{tablenotes}
             \small
			 \item {\bf Notes.} 
			 \item \textsuperscript{a}{These transit depths are calculated for the relevant wavelength ranges using the main +1 order trace.}
            \item \textsuperscript{b}{The bandpasses of the main trace from which these depths are obtained are partly contaminated by flux from the +2 order.}
			\end{tablenotes}
    	\end{center}
	\end{threeparttable}
\end{table*}

\begin{table*}
	\begin{threeparttable}
    	\caption{HST UVIS -1 order Spectroscopic Depths} \label{table:uvis_m1_fits}
    	\begin{center}
        	\begin{tabular*}{\textwidth}{@{\extracolsep{\fill}} c c c c c c c c c c c}
        		\hline \hline
                \multicolumn{5}{c}{-1 order} & 
                \multicolumn{3}{c}{-2 order} & 
                \multicolumn{3}{c}{-3 order}\\
                \cline{1-5} \cline{6-8} \cline{9-11}
        		Bandpass & $\delta_1$ & $\pm 1 \sigma$ & $\delta_{\rm 1, corrected}$ & $\pm 1 \sigma$ &  Bandpass & $\delta_2$\textsuperscript{a} & $\pm 1 \sigma$ & Bandpass & $\delta_3$\textsuperscript{a} & $\pm 1 \sigma$ \\ 
                (nm) & (ppm) & (ppm) & (ppm) & (ppm) & (nm) & (ppm) & (ppm) & (nm) & (ppm) & (ppm) \\ 
                \hline 
                200-220 & 14057 & 317 &  -- & -- &  -- & -- & -- &  -- & -- & --  \\
                220-240 & 14361 & 185 &  -- & -- &  -- & -- & -- &  -- & -- & --  \\
                240-260 & 14687 & 197 &  -- & -- &  -- & -- & -- &  -- & -- & --  \\
                260-280 & 14407 & 174 &  -- & -- &  -- & -- & -- &  -- & -- & --  \\
                280-300 & 14032 & 190 &  -- & -- &  -- & -- & -- &  -- & -- & --  \\
                300-320 & 13798 & 164 &  -- & -- &  -- & -- & -- &  -- & -- & --  \\
                320-340 & 13327 & 145 &  -- & -- &  -- & -- & -- &  -- & -- & --  \\
                340-360 & 13502 & 133 &  -- & -- &  -- & -- & -- &  -- & -- & --  \\
                360-380 & 13573 & 182 &  -- & -- &  -- & -- & -- &  -- & -- & --  \\
                380-400 & 13433 & 125 &  -- & -- &  -- & -- & -- &  -- & -- & --  \\
                400-420 & 13344 & 147 & 13330 & 147 & 189-201 & 14318 & 618 &  -- & -- & --  \\
                420-440 & 13241 & 96 & 13224 & 97 & 201-212 & 13859 & 429 &  -- & -- & --  \\
                440-460 & 13227 & 84 & 13186 & 85 & 212-223 & 14222 & 327 &  -- & -- & --  \\
                460-480 & 13110 & 110 & 13033 & 111 & 223-234 & 14560 & 240 &  -- & -- & --  \\
                480-500 & 13383 & 109 & 13322 & 110 & 234-245 & 14315 & 221 &  -- & -- & --  \\
                500-520 & 13444 & 110 & 13361 & 111 & 245-256 & 14706 & 218 &  -- & -- & --  \\
                520-540 & 13278 & 102 & 13182 & 104 & 256-266 & 14577 & 260 &  -- & -- & --  \\
                540-560 & 13091 & 114 & 12995 & 115 & 266-277 & 14361 & 187 &  -- & -- & --  \\
                560-580 & 13271 & 134 & 13175 & 137 & 277-287 & 14416 & 255 &  -- & -- & --  \\
                580-600 & 13227 & 171 & 13140 & 174 & 287-297 & 14003 & 222 &  -- & -- & --  \\
                600-620 & 13357 & 133 & 13294 & 139 & 297-307 & 13719 & 242 & 194-201 & 14160 & 672 \\
                620-640 & 13484 & 134 & 13382 & 141 & 307-317 & 14069 & 207 & 201-208 & 13710 & 530 \\
                640-660 & 13432 & 169 & 13404 & 179 & 317-327 & 13230 & 189 & 208-215 & 14317 & 479 \\
                660-680 & 13446 & 160 & 13435 & 171 & 327-336 & 13297 & 198 & 215-222 & 13830 & 346 \\
                680-700 & 13512 & 181 & 13407 & 194 & 336-346 & 13387 & 185 & 222-229 & 14500 & 318 \\
                700-720 & 13460 & 192 & 13249 & 209 & 346-356 & 13502 & 188 & 229-236 & 14674 & 273 \\
                720-740 & 13447 & 198 & 13314 & 220 & 356-366 & 13371 & 221 & 236-242 & 14246 & 298 \\
                740-760 & 13419 & 215 & 13130 & 244 & 366-376 & 13655 & 235 & 242-249 & 14410 & 278 \\
                760-780 & 13659 & 260 & 13506 & 309 & 376-386 & 13333 & 200 & 249-256 & 14907 & 274 \\
                780-800 & 13814 & 281 & 13796 & 357 & 386-396 & 13383 & 147 & 256-263 & 14883 & 311 \\
        		\hline
        	\end{tabular*}
            \begin{tablenotes}
             \small
			 \item {\bf Notes.} 
			 \item \textsuperscript{a}{These transit depths are calculated for the relevant wavelength ranges using the main -1 order trace.}
			\end{tablenotes}
    	\end{center}
	\end{threeparttable}
\end{table*}

\section{Results}
\label{sec:results}
\subsection{WLC and limb asymmetries}\label{sec: WLC and Limb}

The best fit astrophysical parameters from the joint white light curve fit are reported in Table~\ref{table:wlc_fits} and the light curves are shown in Figure~\ref{fig:wlcs}. The white light curve depths in the three different bandpasses immediately indicate that planet appears significantly larger in the UVIS bandpass, which indicates the presence of a strongly absorbing species in the $0.2 - 0.8 \, \mu$m wavelength range in the planet's atmosphere. The impact parameter $b$ and relative semi-major axis $a/R_*$ are in excellent agreement (within $1~\sigma$) with published literature values. The mid-transit time of the joint fit is in exquisite agreement with the predicted time for transit based on the ephemeris of \cite{Ivshina2022}. We fix all astrophysical parameters to the best fit values from the joint fit and fit for the mid-transit time for each light curve in our study. The observed minus calculated mid-transit time plot (Figure~\ref{fig:O_C}) shows that the $T_0$ for all our observations agree with the predicted $T_0$.

\begin{figure}
    \centering
    \includegraphics[width=\linewidth]{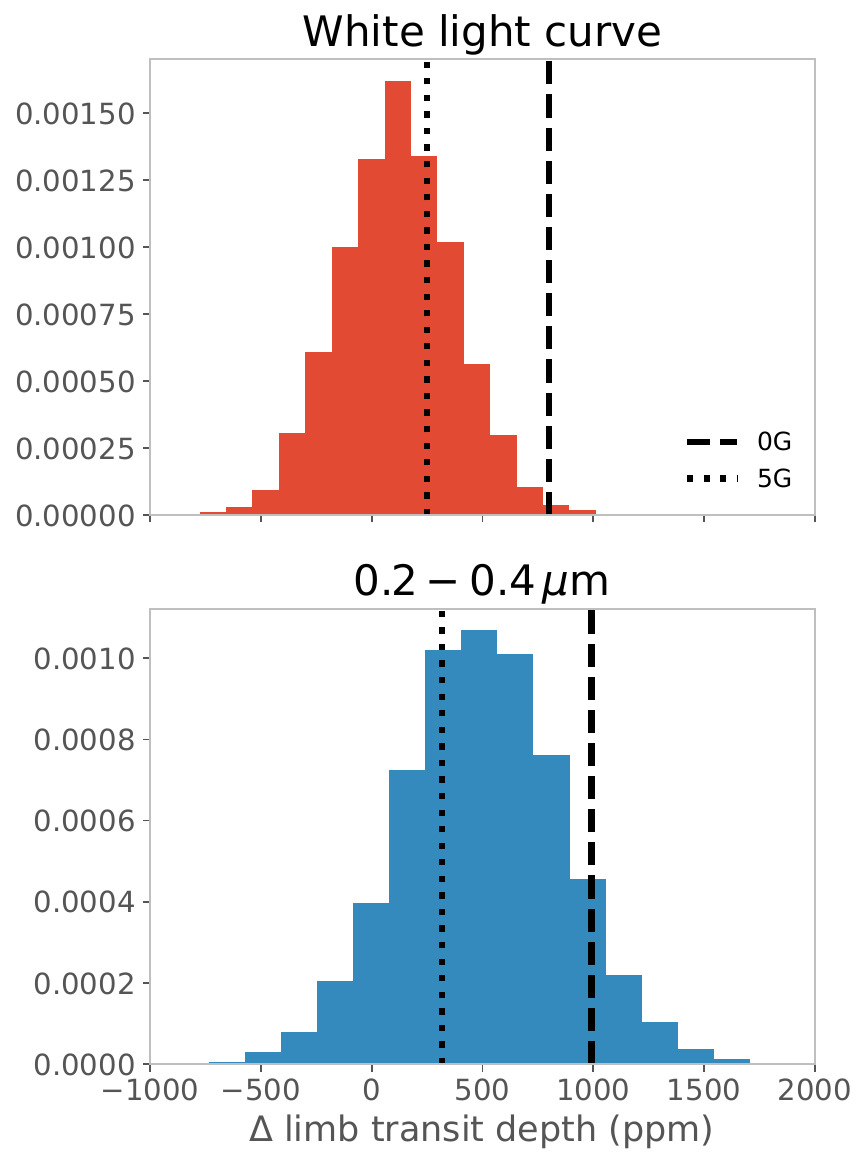}
    \includegraphics[width=\linewidth]{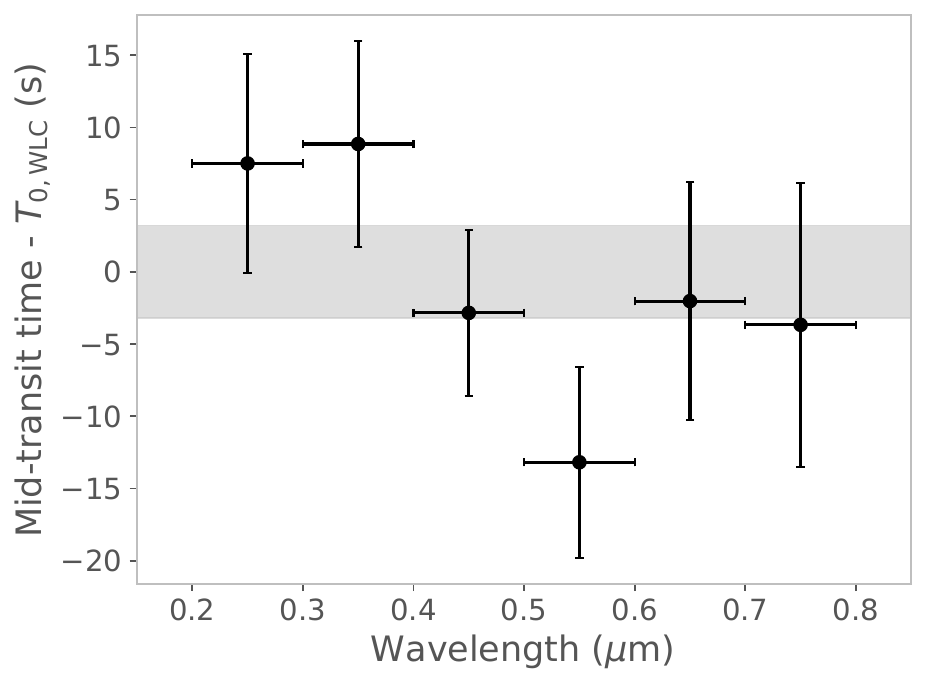}
    \caption{The top and middle panels show the difference in the transit depths for the leading (morning) and the trailing (evening) limbs obtained by fitting the white light curve and the $0.2 - 0.4 \, \mu$m light curve, respectively. The vertical lines show the predicted difference in the transit depths of the two limbs using General Circulation Models (see \S~\ref{sec:gcm}). The bottom panel shows the mid-transit times for light curves binned in $0.1 \, \mu$m-wide bandpasses. These measurements are shown relative to the mid-transit time of the white light curve and the grey region marks the $1~\sigma$ uncertainty of the white light curve's mid-transit time.}
    \label{fig:limb_asymmetry}
\end{figure}

Large horizontal temperature gradients in the atmospheres of tidally locked ultra hot Jupiters are expected to lead to significant asymmetries in the morning and the evening limbs of the planet \citep[e.g.,][]{Wardenier2021, Savel2022}. In general, one expects the evening limb to be hotter, which implies that it has a large scale height and fewer condensed species. Such asymmetries may manifest in a high precision light curve that covers both ingress and egress. The UVIS G280 +1 order light curve is therefore suitable for detecting the signature of any such limb asymmetry. We use \texttt{catwoman} \citep{Jones_catwoman} to fit these data with two separate parameters characterizing the size $R_{\rm p} / R_*$ of the planet on the morning and evening terminators. The angle $\phi$ that determines the orientation of the line separating the two hemispheres with respect to the transit chord is fixed to $90^{\circ}$. The astrophysical parameters $a/R_*$, $b$, and $T_0$ are fixed to best fit values from the joint white light curve fit. The quadratic limb darkening parameters and the systematics model parameters are allowed to vary.

Figure~\ref{fig:limb_asymmetry} (top panel) shows the constraints on the transit depths for the two limbs obtained from fitting the full +1 order white light curve. Over this wide wavelength range, there is no hint of an asymmetry between the morning and the evening limbs ($118 \pm 258$ ppm). The wavelength region where the largest difference between the two limbs might be expected is $0.2 - 0.4 \, \mu$m, where we observe the stupendous rise in transit depths that corresponds to $> 10$ atmospheric scale heights. In this wavelength region, the transit depths for the two limbs differ only at the $1.4~\sigma$ level ($500 \pm 360$ ppm), with the evening limb being larger than the morning limb as expected. We also fit for the mid-transit time of the +1 order light curves split into 100 nm bins to test if there is a significant variation in $T_0$ with wavelength as a result of limb asymmetries. In Figure~\ref{fig:limb_asymmetry} (bottom panel), we show these values in comparison to the best fit $T_0$ from the joint white light curve fit. These measurements indicate that there is tentative evidence of the mid-transit occurring a little later at shorter wavelengths, which is compatible with the trailing (evening) limb being larger than the leading limb. Overall though, our UVIS +1 order light curve hints at the absence of detectable limb asymmetries. However, more observations are needed to firmly establish this result.

\subsection{Transmission Spectrum of KELT-20 b} \label{sec:spectrum}

\begin{figure*}
    \includegraphics[width=\linewidth]{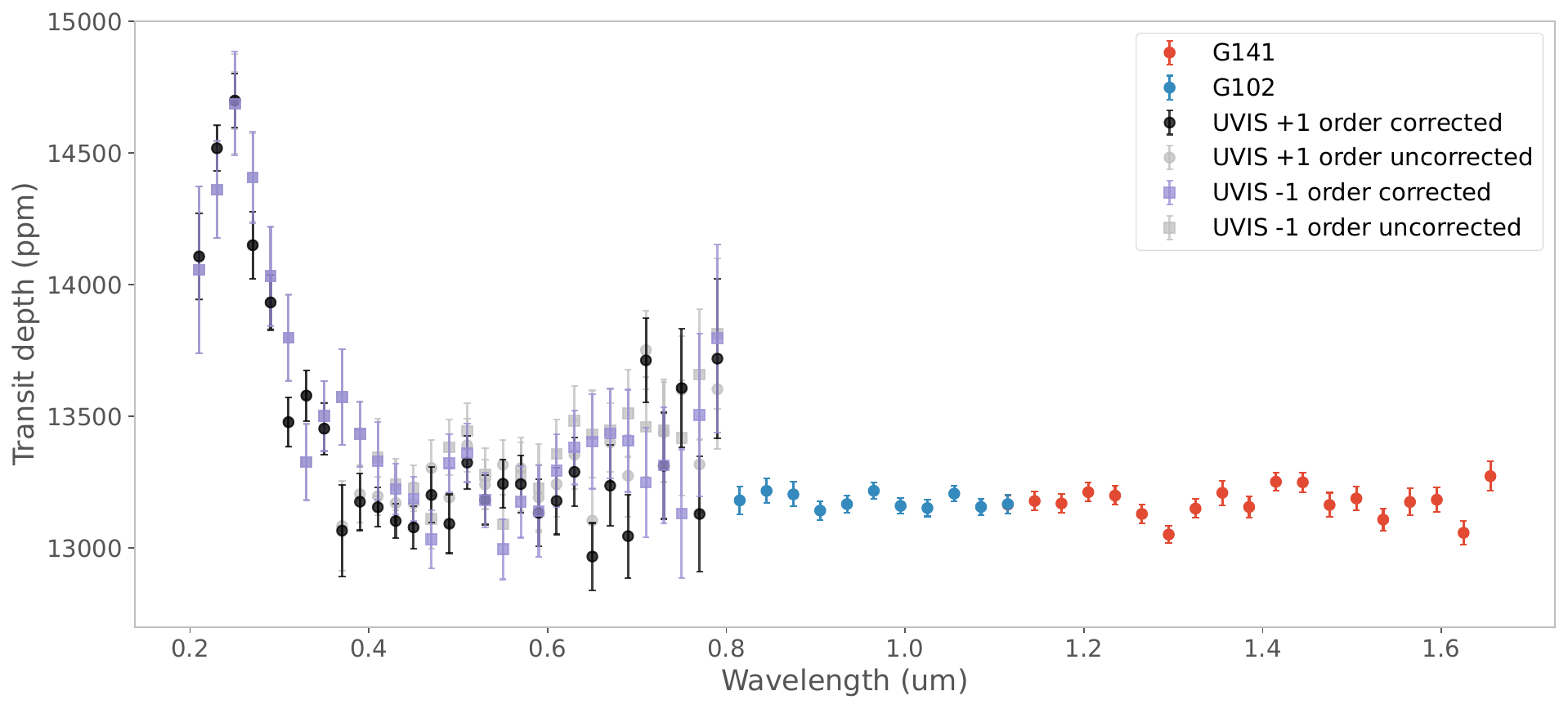}
    \caption{The $0.2 - 1.7 \, \mu$m transmission spectrum of KELT-20 b measured with the WFC3 instrument in the G280 (UVIS), G102, and G141 bandpasses.}
    \label{fig:full_spectrum}
\end{figure*}

The spectroscopic transit depths are listed in Table~\ref{table:g102_g141_fits}, \ref{table:uvis_p1_fits}, and \ref{table:uvis_m1_fits}. For the UVIS spectroscopic depths, we list the fitted transit depths of the planet for the wavelength range corresponding to the contaminating order as well. These depths are used to correct for the effect of order overlap (see \S~\ref{sec:extraction} for details). Figure~\ref{fig:full_spectrum} shows the transmission spectrum of KELT-20 b with both uncorrected and corrected UVIS depths and the G102 and G141 depths. There is a startling rise in the transit depth at wavelengths shorter than $0.4 \, \mu$m where the apparent size of the planet grows by nearly 6\%. The two independently reduced UVIS spectra from orders +1 and -1 are in reasonably good agreement. The largest discrepancy is in the $360-380$ nm bandpass where the stellar flux drops precipitously due to ionization of hydrogen from its first excited state ($n = 2 \rightarrow \infty$, `Balmer jump') and small sub-pixel shifts in the spectrum have a strong impact on the transit light curve in this wavelength range. Redward of $0.6 \, \mu$m, contamination from higher orders has a significant effect on the measured transit depths and their precision.

The NIR transit depths sample a much smaller range of pressures/atmospheric scale heights. The G102 spectrum is nearly flat. In the overlapping bin ($1.13 - 1.16 \, \mu$m) between the G102 and G141 bandpasses, the transit depths agree to within $1.4 \, \sigma$ and suggest a negligible offset between the two bandpasses. The G141 transit depths display two broad shallow features: the one centered at $1.4 \, \mu$m hints at absorption due to water. The feature at $1.1-1.3 \, \mu$m could potentially be due to FeH \citep{Evans2018} but its origin remains ambiguous as it is not well-matched by any known molecules with opacities in our databases (\texttt{PLATON} \citealt{Zhang2018}, \texttt{PETRA} \citealt{Lothringer2020_petra}, \texttt{petitRADTRANS} \citealt{molliere_2019}). We note that both features in the G141 spectrum are relatively weak and the Bayes factor ($\mathcal{B} = 12.9$) mildly favors fitting the spectrum with a flat line rather than with a flat line + Gaussian model.

\section{Retrieval Analysis}
\label{sec:retrievals}

\subsection{PETRA}
To quantify the constraints our observations place on the atmospheric properties of KELT-20 b, including the temperature structure and composition, we perform a retrieval analysis with the PHOENIX Exoplanet Retrieval Algorithm, or \texttt{PETRA} \citep{Lothringer2020_petra}. \texttt{PETRA} uses the PHOENIX 1D atmosphere model \citep{Hauschildt1999} as its forward model within a differential evolution MCMC statistical framework \citep{TerBraak2006, TerBraak2008}. We run the retrieval with 30 chains for a total of 45000 iterations until the Gelman-Rubin statistic is below 1.1, trimming the first 500 iterations of each chain as burn-in. 

We use the 3-parameter temperature profile parametrization from \cite{Guillot2010}, which allows for a temperature inversion. We fit for an offset between G280 and G102+G141 due to the lack of reliable overlap. We fix the surface gravity to $\log{g} = 3.3$, which corresponds to a mass of about 2.4~$M_J$. Analyses were run both in chemical equilibrium and as a so-called free retrieval, where the abundance of individual molecules, atoms, and ions are allowed to freely vary with a vertically-uniform abundance. In chemical equilibrium, we retrieve both a [O/H] and [Z/H] metallicity abundance ratio, relative to solar, the latter of which represents all heavy elements besides oxygen. 

In the free retrieval, we retrieve the abundance of H$_2$O, TiO, SiO, Fe I, Fe II, Mg I, Mg II, and e$^-$ (which determines the abundance of H$^-$). The abundance of remaining species in the background atmosphere (e.g., Ni, Ti, Al, Ca, VO, FeH, etc.) were included in chemical equilibrium, parameterized through an additional [Z/H] metallicity free parameter.

\subsubsection{Chemical Equilibrium Results}

\begin{figure*}
    \centering
    \includegraphics[width=\linewidth]{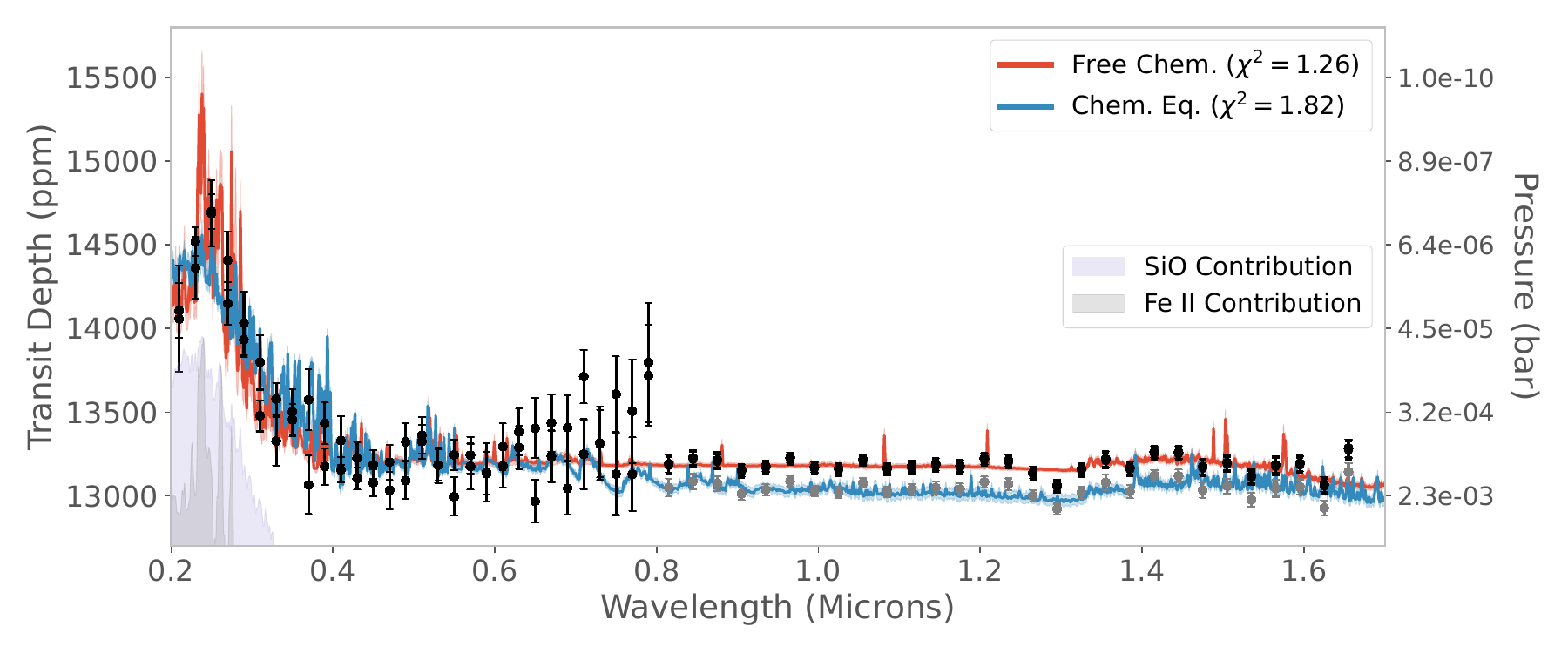}
    \caption{Median fit spectra with the 1-$\sigma$ range from the free chemistry (red) and chemical equilibrium (blue) \texttt{PETRA} retrievals. In the calculation of the reduced-$\chi^2$ value, the best-fit offset of +10.7 and $-130.1$ ppm was applied to data points longwards of 0.8 $\mu$m to the free chemistry and chemical equilibrium retrievals, respectively. This offset is shown by the gray points. The purple and gray shaded regions at short wavelengths show the contribution to the transit depth from 10 ppm of SiO and the best-fit abundance of Fe II, respectively. The approximate pressure is listed on the right axis, though the relationship between transit depth and pressure varies sample-to-sample as a function of temperature and atmospheric opacity.}
    \label{fig:petra_spec_fits}
\end{figure*}

The chemical equilibrium retrieval qualitatively fits the 0.2-1.7~$\mu$m transit spectrum of KELT-20 b (see Figure~\ref{fig:petra_spec_fits}), but with $\chi^2/N_{data} = 1.82$, the quantitative goodness-of-fit implies either a model inadequacy or underestimated error bars. In the event of underestimated error bars, the corresponding uncertainty on the constraints on the planet's atmosphere will be underestimated. For the WFC3-IR data alone (0.8-1.7$\mu$m), $\chi^2/N_{data} = 1.64$ and for the UVIS data (0.2-0.8$\mu$m), $\chi^2/N_{data} = 1.91$, indicating that the fit to the NUV-optical data is somewhat worse than the fit to the IR data.

\begin{figure}
    \centering
    \includegraphics[width=\linewidth]{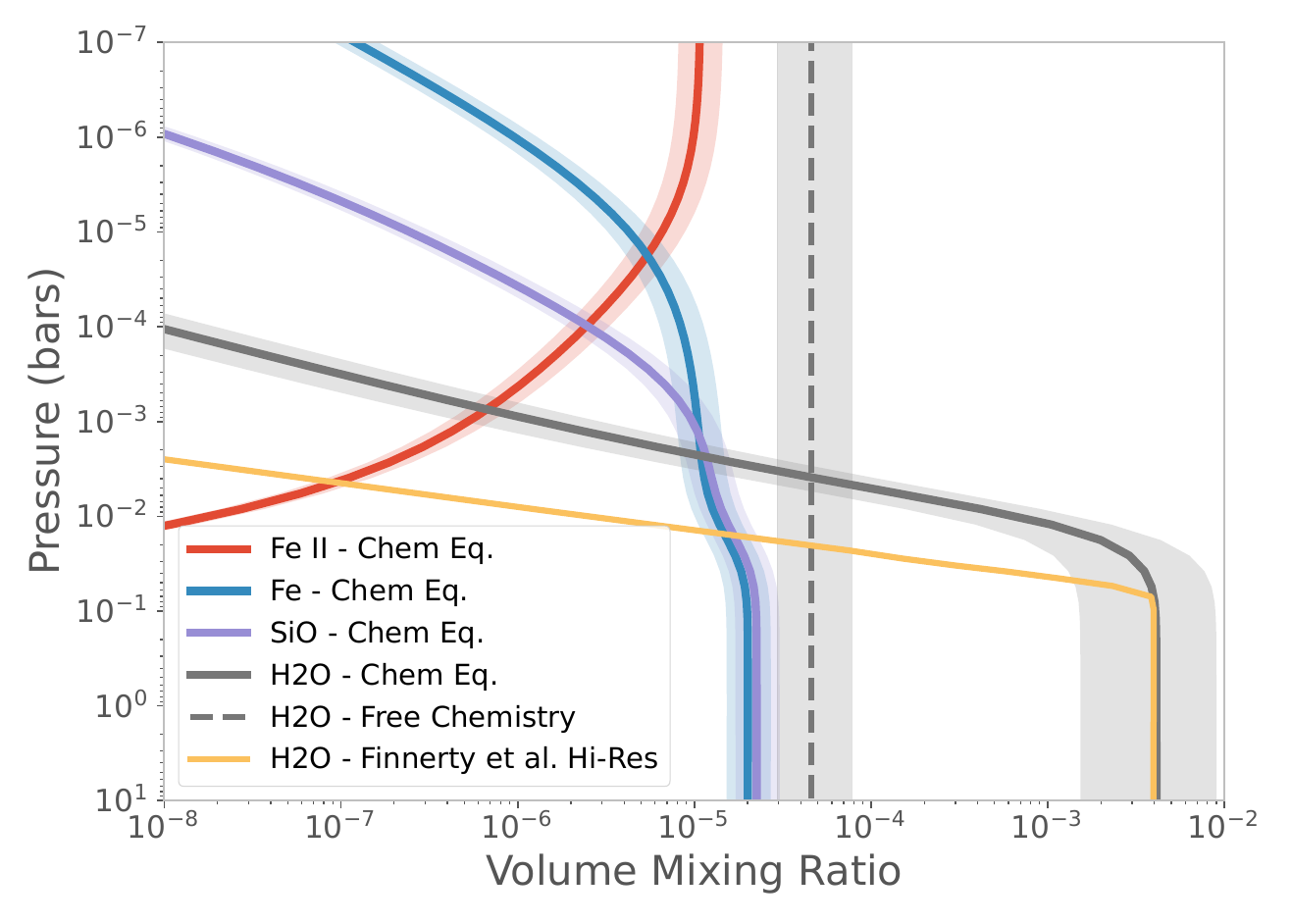}
    \caption{Comparison of the volume mixing ratio and 1-$\sigma$ range of Fe II, Fe I, SiO, and H$_2$O from the chemical equilibrium retrieval with the H$_2$O abundance from the free chemistry retrieval and the median-fit from \citep{Finnerty2025}.}
    \label{fig:h2o_abund_compare}
\end{figure}

\begin{figure}
    \centering
    \includegraphics[width=\linewidth]{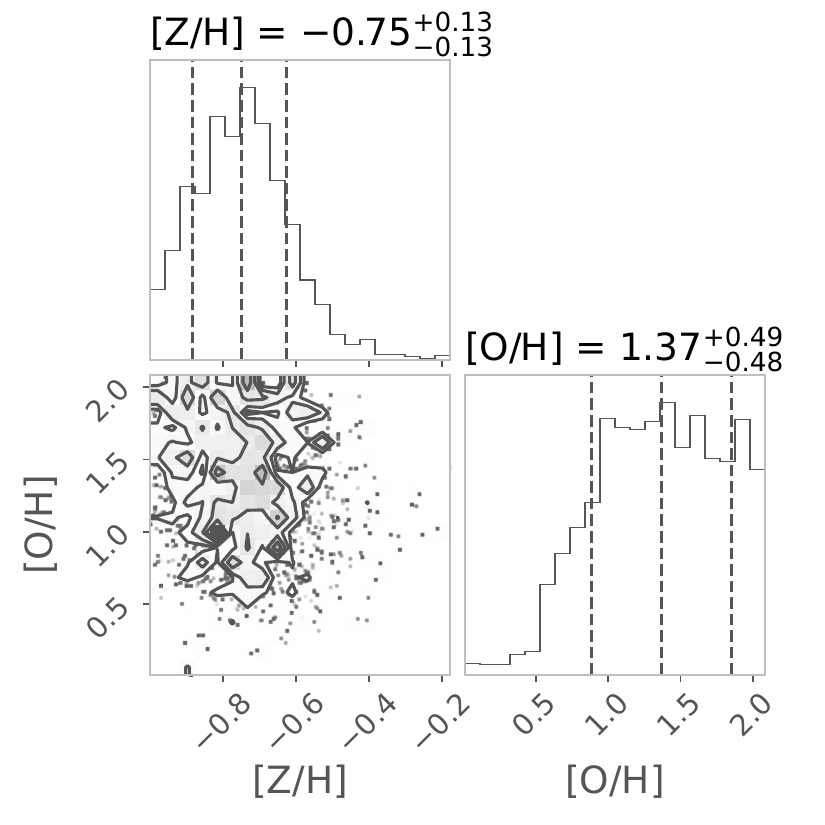}
    \caption{[Z/H] and [O/H] ratio posterior constraints from the chemical equilibrium retrieval with \texttt{PETRA}.}
    \label{fig:petra_eq_corner}
\end{figure}

Figure~\ref{fig:h2o_abund_compare} shows the resulting abundances of Fe I, Fe II, SiO, and H$_2$O from the best-fit retrieved chemical equilibrium model. [O/H] is constrained to be greater than 7.7$\times$ solar abundance at 1-$\sigma$, with the upper bound running into our upper-limit uniform prior at 2.0 (see Figure~\ref{fig:petra_eq_corner}). This abundance of H$_2$O is consistent with the retrieved constraints from high-resolution KELT-20 b observations, also shown in Figure~\ref{fig:h2o_abund_compare}, though \cite{Finnerty2025} measure the dissociation to begin at a slightly higher pressure. The [O/H] ratio is constrained by both the presence of H$_2$O and SiO, though [Z/H] plays a role in setting the limiting Si abundance. Interestingly, [Z/H] is found to be subsolar at $-0.75 \pm 0.13$.

The major complication with both constraining and interpreting the [O/H] abundance is thermal dissociation; at the ultra hot temperatures of KELT-20 b ($\approx2500$~K), H$_2$O dissociates into H and OH. While this is taken into account through the coupling of the equilibrium chemistry and the temperature structure, it means that the H$_2$O abundance at the transit photosphere that we are sensing with our observations is relatively low, even though the global O/H is greater than 10$\times$ solar (compatible with O/H measured in emission, \citealt{Fu2022}), as seen in Figure~\ref{fig:h2o_abund_compare}. The H$_2$O volume mixing ratio preferred by the retrieval is closer to $10^{-5}$, which is less than the abundance of H$_2$O under more typical (i.e., slightly cooler) hot Jupiter conditions.

Changes in the temperature structure or the presence of photo-dissociation (which is not currently taken into account in the retrieval) can modify the bulk [O/H] abundance inferred from the H$_2$O spectral features. The constraints on the temperature structure from these transit observations is limited (Figure~\ref{fig:tpprofiles}), but the chemical equilibrium retrieval prefers a fairly isothermal structure at about 3350~K. While a temperature inversion may still be present at the terminator, the high temperature may also be driven by the need to create such a large NUV transit feature. The higher temperature leads to a larger scale height, and thus a larger NUV spectral feature. This could then be related to the super-solar [O/H] constraint, whereby the retrieval increases the [O/H] abundance to generate enough non-dissociated H$_2$O to match the spectral feature at 1.4~$\mu$m. 

\subsubsection{Free-Retrieval Results}
We also ran a free-chemistry retrieval, which allows for greater flexibility in fitting the spectrum, as chemical abundances may deviate from equilibrium. Overall, the free chemistry retrievals do indeed fit much better at all wavelengths than the chemical equilibrium retrieval, finding a full-wavelength $\chi^2/N_{data}$ of 1.26, with a WFC3-IR-specific $\chi^2/N_{data}$ of 1.45 and UVIS-specific $\chi^2/N_{data}$ of 1.17. Thus, the free chemistry retrieval fits the UV-optical observations better than the IR data, opposite to the behavior of the chemical equilibrium retrieval. Unlike the chemical equilibrium retrieval, the free retrieval has a high degree of flexibility for fitting the short-wavelength observations with a variety of different absorbers.

\begin{figure*}
    \centering
    \includegraphics[width=\linewidth]{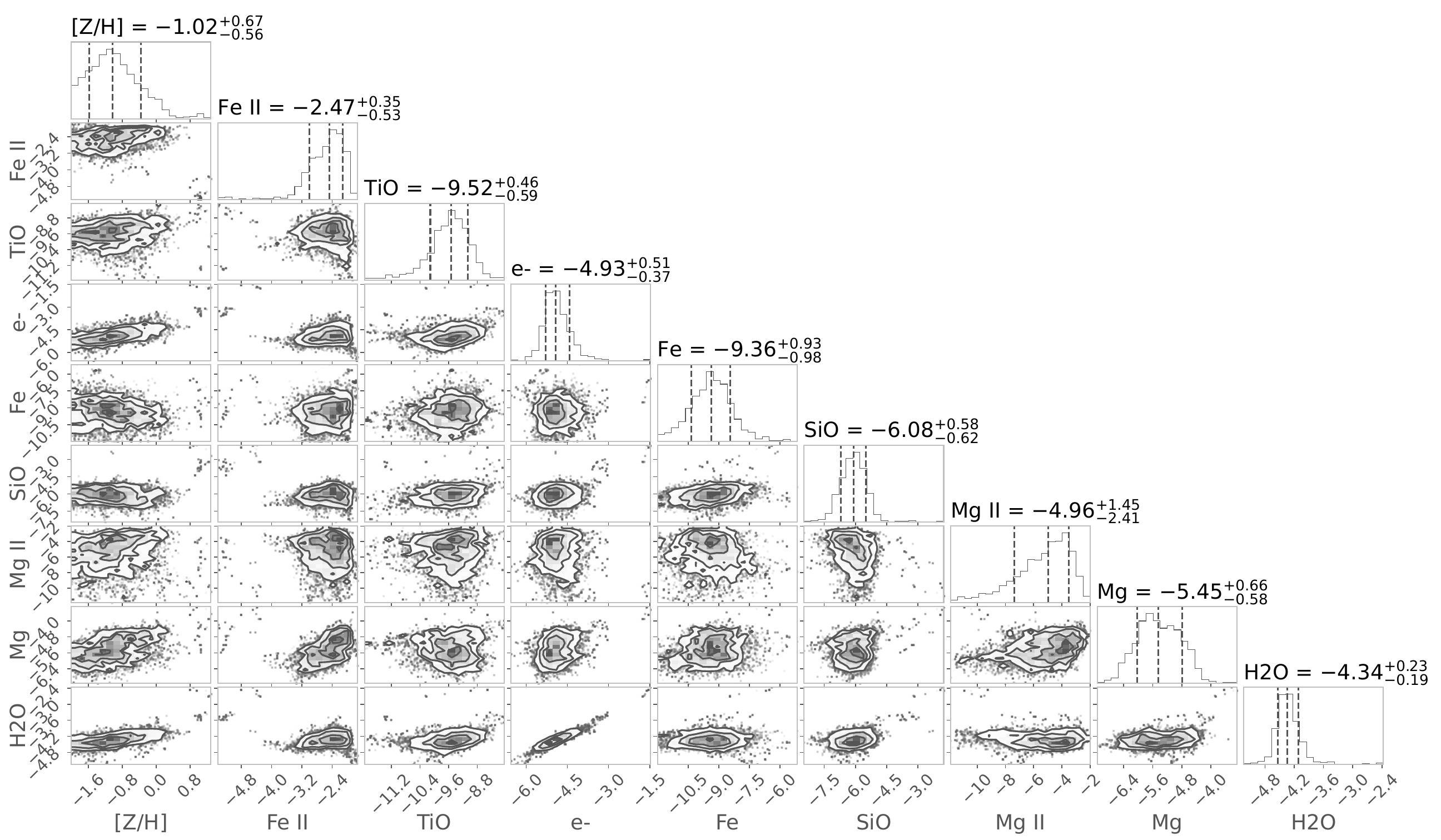}
    \caption{Abundance posterior distributions for the free chemistry retrieval with \texttt{PETRA}.}
    \label{fig:petra_free_corner}
\end{figure*}

However, the abundance constraints are different from the chemical equilibrium retrieval (see Figure~\ref{fig:petra_free_corner}). While a mixture of SiO and Fe II contribute to the fit in chemical equilibrium, in the free retrieval, Fe II is the dominant opacity source in the NUV, helping to match the shape of the spectral feature, which reaches a maximum at approximately 0.23-0.25 $\mu$m. The retrieved Fe II abundance is, however, unphysical at a log volume mixing ratio of $-2.47^{+0.35}_{-0.53}$, which corresponds to $\sim 100\times$ solar. While this high Fe II abundance may be due to an incorrect attribution of the feature to Fe II by the retrieval, 
a more plausible Fe II abundance could potentially be obtained by dropping the LTE approximation in the middle and upper atmospheric regions \citep{Fossati2023}. Accounting for NLTE effects increases the temperature in the middle atmosphere, which in turn strengthens the NUV FeII lines and reduces the abundance of Fe II required to match the observations. These effects can also reduce the amount of SiO required to match the NUV observations \citep{Fossati2025}.. 

Although an escaping or non-hydrostatic atmosphere could also enhance Fe II column density higher up in the atmosphere, the effect of an outflow on the density profile (density enhancement $\propto e^{-u^2/2c_{\rm s}^2}$ for an isothermal wind, where $u$ is the wind speed and $c_{\rm s}$ is the sound speed) in the sub-sonic region ($u < c_{\rm s}$) of an atmosphere is negligible \citep{Parker1958}.

Additionally, we note that Fe I only has an upper-limit constraint, implying that most of the Fe appears to be ionized at the transit photosphere of KELT-20 b. Other species are also constrained, including Mg I, Mg II, H$^-$ (through e$^-$), and TiO, though their unique contributions can be quite minor. For example, a retrieval without Mg I found an equally good fit with a $\Delta$BIC of $-4.5$, implying Mg I is not necessary to fit the data. On the other hand, including Fe II greatly improves the fit with a $\Delta$BIC of 13.9 indicating that Fe II's inclusion is necessary to fit the data. The case of TiO is also interesting, because the optical data where TiO is best constrained does not appear to be at high enough precision to uniquely identify TiO features, and thus we interpret TiO as the species setting the optical baseline in the retrieval.

Finally, H$_2$O's log mixing ratio is measured to be $-4.34^{+0.23}_{-0.19}$, in line with the low abundance seen in Figure~\ref{fig:h2o_abund_compare} for both the chemical equilibrium retrieval and the high-resolution retrieval of \citep{Finnerty2025}. To reiterate, our retrievals for this planet appear to be retrieving the highly dissociated H$_2$O abundance at about 1-10 millibar, making interpretation of the great bulk oxygen abundance difficult. However, it is worth noting that all three KELT-20 b retrievals show roughly similar H$_2$O abundances at 1-10 millibar. 

\subsubsection{Condensation Temperature Rainout Retrieval}
\label{sec:t_cond_retrieval}
We also performed a retrieval where the abundance of atmospheric species were in chemical equilibrium, but a free-parameter $T_\mathrm{cond}$ determined the condensation/rainout temperature of the atmosphere. Elements with condensation temperatures above $T_\mathrm{cond}$ (i.e., more refractory), are depleted by 3 orders of magnitude before the calculation of chemical equilibrium. We use the elemental condensation temperatures of \cite{Wood2019}. Since $T_\mathrm{cond}$ is a free-parameter and is completely independent of the temperature structure, this scenario simulates either vertical or nightside rainout quenching.

This scenario retrieved a very similar atmosphere to the original chemical equilibrium retrieval, with $T_\mathrm{cond}$ constrained to be above 1565~K, which corresponds to the condensation temperature of Ti into perovskite, CaTiO$_3$. The only species that have condensed at $T_\mathrm{cond}$, and are thus depleted in the retrieved scenario, are Al (at 1652~K into corundum, Al$_2$O$_3$) and Zr (at 1722~K into ZrO$_2$). The reason the retrieval needs Ti to remain uncondensed (i.e., undepleted), is because TiO is setting the optical baseline continuum. Without TiO, the optical transit depths would be much shallower. We explore an alternative avenue towards constraining the condensation temperature by comparing to the abundances retrieved from high-resolution spectra in Section~\ref{sec:disc:hires}, finding a similar lower limit to $T_\mathrm{cond}$.

\subsection{petitRADTRANS}
\label{sec:petit_retrievals}

We perform a secondary retrieval analysis using the open-sourced retrieval framework \texttt{petitRADTRANs} \citep{molliere_2019}. Similar to the \texttt{PETRA} retrievals, we used the temperature profile parametrization from \cite{Guillot2010}. We fit for the surface gravity, log $g$, and the planet radius at the reference pressure 0.01 bars, $R_p$. We perform both free chemistry retrievals, where the chemical abundance is fixed with altitude, and equilibrium chemistry retrievals. Our equilibrium chemistry retrievals are parameterized by metallicity, scaled relative to the solar iron abundance, $[$Fe/H$]$, and $[$O/H$]$ ratio, again normalized to the solar value, which sets the oxygen abundance. We calculate the equilibrium chemical abundances based on these two parameters using the equilibrium chemistry code easyCHEM, a CEA \citep{gordon_1994,mcbride_1996} clone first described in \cite{molliere_2017}. For both retrievals, we include opacities for H$_2$O, FeH, TiO, SiO, VO, Fe, Fe II, Mg, Mg II, Ti, Ti II, Na and K. We modify the correlated-k opacities included with \texttt{petitRADTRANS} to extend to 0.1 microns for Fe II and SiO to model our UVIS spectrum. We include continuum opacities from collision-induced absorption from H$_2$/H$_2$,  H$_2$/He, and H$^-$. For our free retrievals, we included H$^-$, H, and e$^-$ as free parameters, and calculated their abundances using easyCHEM in our equilibrium chemistry retrievals. We additionally fit for offsets between each of the instrument modes. Our priors and retrieved posteriors for these fits are shown in Table~\ref{tab:prt_retrieval_comparison}.

\subsubsection{Equilibrium Chemistry Retrievals}
Similar to the \texttt{PETRA} equilibrium chemistry retrieval results, we find a subsolar metallicity, $[$Z/H$]$~=~$-1.25^{+0.18}_{-0.15}$, which is $2.3~\sigma$ lower than the metallicity retrieved by \texttt{PETRA}. We also retrieve a similar enhancement of oxygen, with the best fit value for $[$O/H$]$~=~$1.33^{+0.45}_{-0.82}$. Unlike the \texttt{PETRA} retrievals, we fit for the surface gravity (log $g$)and retrieve a best fit value of $3.1\pm0.1$, which equals a best-fit mass of $1.9\pm0.4$ M$_{J}$. We observe strong covariances between the temperature, log $g$ and $[$O/H$]$, which are likely responsible for the broader constraints relative to the \texttt{PETRA} equilibrium retrievals. In particular, the largest values of $[$O/H$]$ correspond to lower surface gravity. The difference in the assumed planet mass as well as differences in opacities and the species included in the retrievals are likely responsible for the quantitative differences in the retrieved abundances.

We find that our chemical equilibrium retrievals are decisively preferred by a ln-Bayes factor of 16.7. This preference is likely due to the fewer degrees of freedom in the retrieval.

\subsubsection{Free Chemistry Retrievals}
We perform multiple free retrievals using the above set up to compare to the \texttt{PETRA} free retrievals. We run retrievals including both SiO and Fe II, and removing one or the other. We find that the retrieval in which SiO contributes to the large NUV feature is strongly preferred over retrievals where SiO is removed and Fe II is the dominant source of NUV opacity, with a Bayes factor of 11. This requires an SiO abundance of $-3.7\pm0.2$ (VMR). In this retrieval, SiO, H$_2$O and K are the only species with bound abundances, though H$_2$O ($-4.1^{+1.2}_{-4.0}$ VMR) has a long tail to low abundances, indicating there are equivalently good fits where they do not contribute much to the opacity. In our retrievals where SiO is removed, we instead find a best fit Fe II abundance of $-2.7\pm0.3$ (VMR). In this retrieval, we also find a bound abundance of TiO ($-10.2\pm0.2$ VMR) and Fe ($-8.1\pm1.0$ VMR). H$_2$O ($-5.6\pm0.2$ VMR) has a much narrow posterior than in our retrieval that includes SiO, though the peak values of both posteriors are still consistent. K has a substantially lower abundance in the retrieval without SiO than when SiO is present. K has several strong lines between 300 and 400 nm, at the edge of the strong NUV feature, so its abundance could be trading off with other species in order to reproduce this feature. The strong K doublet at $\sim766$ nm and $\sim770$ nm lie in a noisy part of the spectrum, which masks any obvious K feature and is likely also driving this degeneracy.  

\section{Comparison with other ultra hot Jupiters}

\begin{figure*}
    \includegraphics[width=\linewidth]{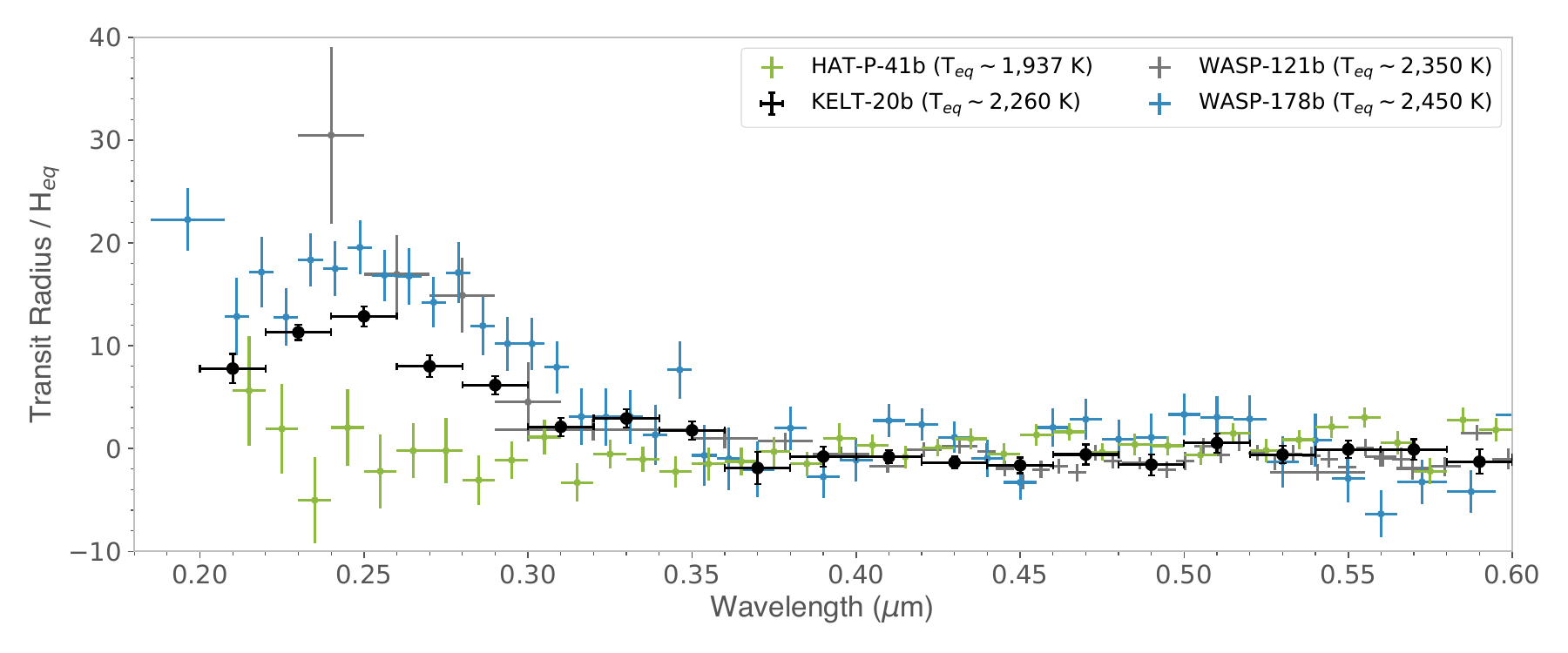}
    \includegraphics[width=\linewidth]{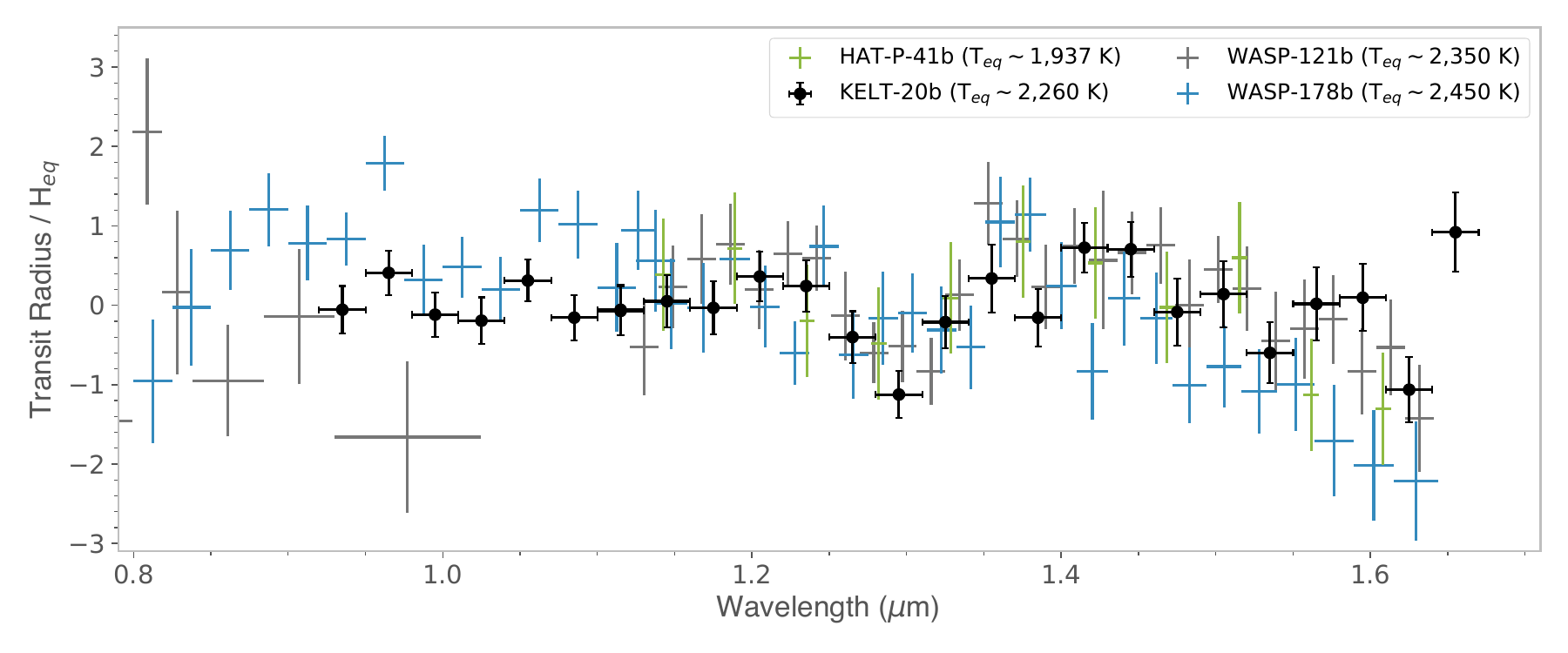}
    \caption{Comparison of KELT-20 b's NUV-NIR spectrum with published transit spectra for HAT-P-41 b \citep{Wakeford2020}, WASP-178 b \citep[][Lothringer et al., in prep]{Lothringer2022}, and WASP-121 b (STIS/E230M, \citealt{Sing2019}). The transit spectra are normalized to their equilibrium scale height (assuming a mean molecular weight of 2.3). For KELT-20 b, we assume a mass of 2 M$_{\rm Jup}$ to calculate the surface gravity for the scale height. While HAT-P-41 b is conspicuously featureless at NUV wavelengths, all three hotter planets show strong NUV absorption, extending over 10 scale heights.}
    \label{fig:spec_comparison}
\end{figure*}

\begin{figure*}
    \includegraphics[width=0.5\linewidth]{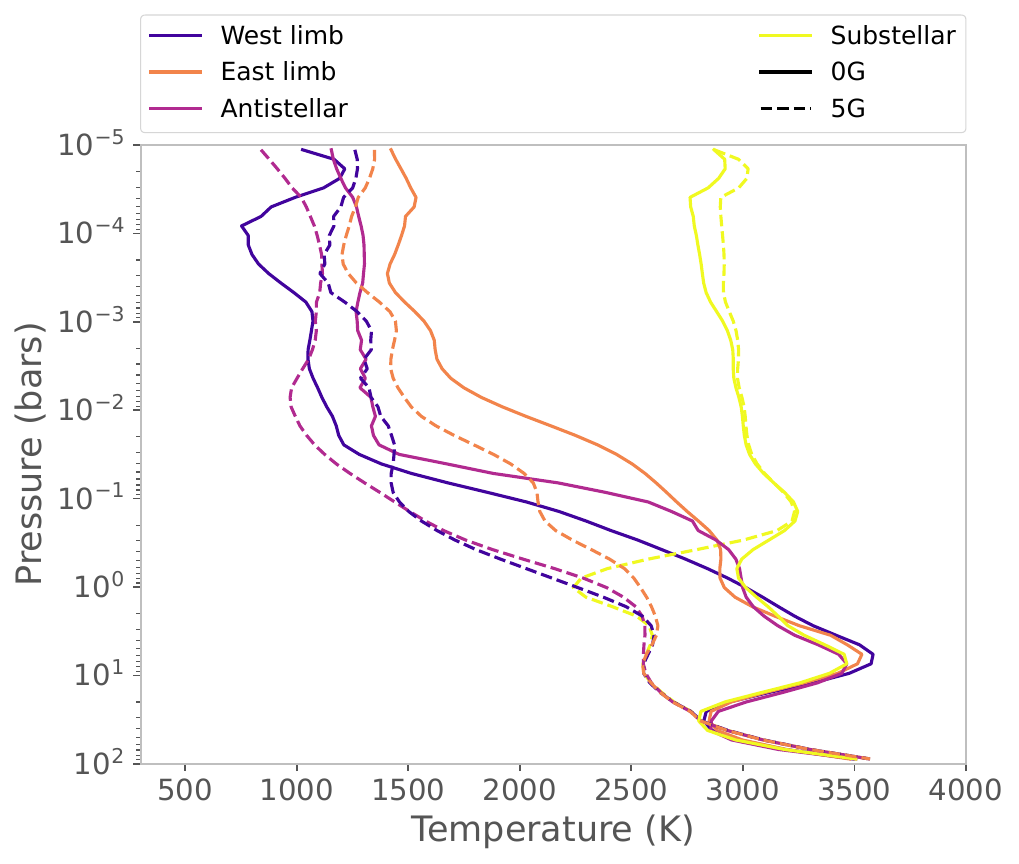}
    \includegraphics[width=0.5\linewidth]{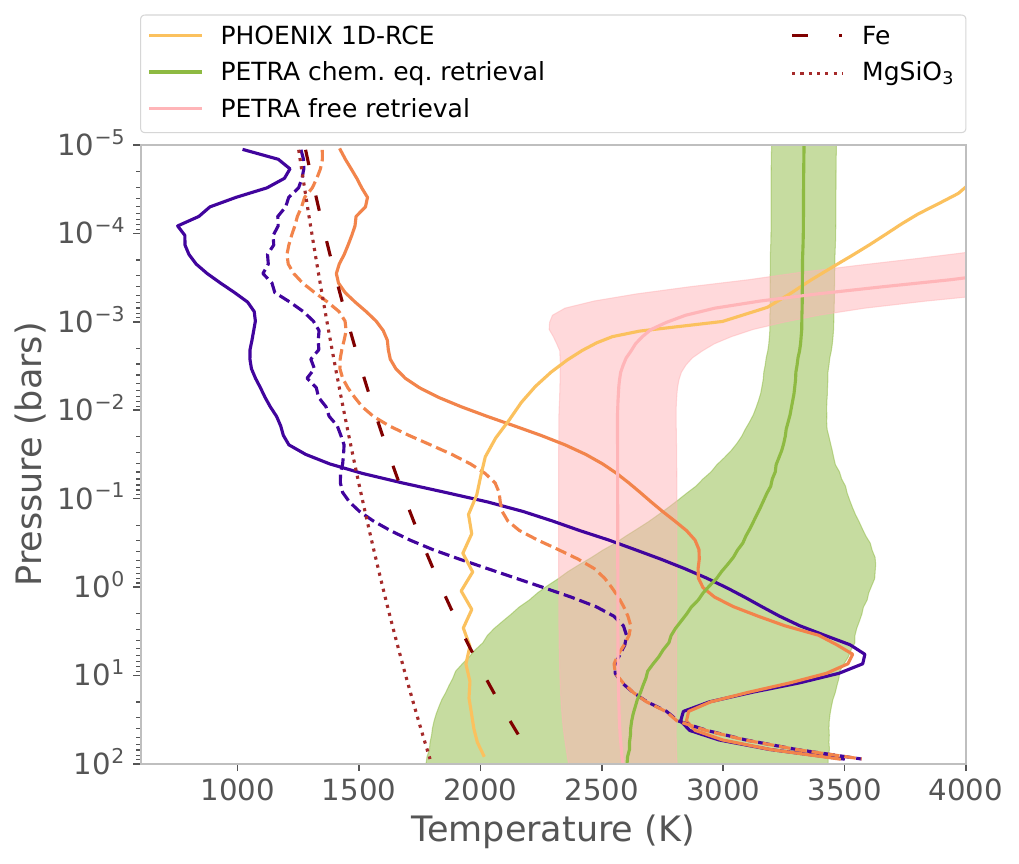}
    \caption{Left: Time averaged temperature-pressure profiles for four regions of  the 0G/drag-free GCM (solid lines) and the 5G active drag GCM (dashed lines). We can see that the active drag model has a slightly hotter dayside and a significantly cooler nightside. We also note that the east limb is hotter in the drag free case due to more efficient heat transfer on the dayside. The western limb is hotter in the active drag case in the upper atmosphere due to differences in the flow patterns between the two models.
    Right: The GCM limb profiles are plotted along with the retrieved temperature-pressure profiles from \texttt{PETRA}, the 1D radiative-convective equilibrium (1D-RCE) model from \texttt{PHOENIX} assuming full heat redistribution, and the condensation curves for Fe and MgSiO$_{3}$ from \cite{Visscher2010}.}
    \label{fig:tpprofiles}
\end{figure*}

In Figure~\ref{fig:spec_comparison}, we compare the NUV and NIR spectrum of KELT-20 b with the spectra of a sample of  UHJs observed with \emph{HST}. At $T_\mathrm{eq}\sim$2250 K, KELT-20 b helps fill a gap in equilibrium temperature compared to other gas giants observed with HST/WFC3-UVIS/G280. The NUV-optical spectrum of HAT-P-41 b ($T_\mathrm{eq} \sim $1950 K) is relatively featureless \citep{Wakeford2020}, while WASP-178 b ($T_\mathrm{eq} \sim $2450 K) exhibits an enormous NUV feature \citep{Lothringer2022}. When combined with the binned-down HST/STIS NUV and optical spectra of WASP-121 b ($T_\mathrm{eq} \sim $2350 K, \citealt{Sing2019}), a clear trend emerges: somewhere between the equilibrium temperatures of HAT-P-41 b and KELT-20 b there appears to be a transition from the cloudy, featureless NUV spectra of hot Jupiters to the large NUV refractory absorption typical of ultra hot Jupiters. This NUV feature is likely due to absorption from a combination of SiO, Fe II and Mg (as we find for KELT-20 b and as suggested for WASP-178 b, \citealt{Lothringer2022}). Given the similarity of the condensation temperatures of iron and silicate species in planetary atmospheres (Figure~\ref{fig:tpprofiles}, right panel), which of these species dominates absorption in the NUV should not significantly limit our ability to interpret this transition.

1D models \citep[e.g.,][]{Visscher2010, Gao2020} predict that the iron and silicate species should first begin to appear in gas phase for hot Jupiters with equilibrium temperatures between $1700-2200$~K. The exact onset of the cloud condensation is sensitive to the precise particle microphysics, as well as the state of the atmosphere (e.g., temperature, mixing, metallicity). In addition, for the clouds to be visible on the limbs, nightside temperatures (which depend on atmospheric circulation patterns set by $T_{\rm eq}$, metallicity, rotation, surface gravity, magnetic field, etc.) must be warm enough to prevent cold-trapping of the condensate species (see constraints on $T_{\rm cond}$ for KELT-20 b in \S~\ref{sec:t_cond_retrieval} and \ref{sec:disc:hires}). The absence or presence of gaseous refractory species as seen in NUV transit spectra, like the one we present here, can therefore signal whether such clouds have condensed or not, respectively \citep{Lothringer2020b}.

While we have assumed a mass of 2 M$_{\rm Jup}$ for KELT-20 b for the purposes of Figure~\ref{fig:spec_comparison}, uncertainty in the mass leads to uncertainty in the surface gravity and therefore the scale height used to normalize and compare the spectra. A higher (lower) mass for KELT-20 b would decrease (increase) the scale height and therefore increase (decrease) the number of scale heights over which we observe the NUV spectrum. If KELT-20 b is indeed 2 M$_{\rm Jup}$ or less however, then the NUV transit depths appear to be somewhat shallower than those seen in WASP-121 b or WASP-178 b. 
This might suggest that this condensation-driven transition is not `sharp' and partial rainout of refractories on the nightside leads to intermediate refractory abundances and NUV feature heights (similar to what is observed for hot Jupiters, \citealt{Sing2016}). However, our constraints on nightside condensation via $T_{\rm cond}$ (\S~\ref{sec:t_cond_retrieval} and \ref{sec:disc:hires}) indicate that only the most refractory species (e.g., Al, Zr, and potentially Ti) may be condensed out. Alternatively, the lower feature height for KELT-20 b could be due to an overall lower abundance of the absorbing refractory species. A measurement of KELT-20 b's mass, more HST/WFC3/G280 transit spectra of hot Jupiters in this temperature range and measurement of their refractory abundances would better resolve the hot-to-ultra hot Jupiter transition and reveal any possible intrinsic diversity or trends with other planet properties (e.g., surface gravity).

The NIR spectra of the four UHJs are shown in the bottom panel of Figure~\ref{fig:spec_comparison}. The water feature at 1.4~$\mu$m appear similar for these planets. In the $1.1 - 1.3~\mu$m region, WASP-121 b displays a shallow feature that \cite{Evans2018} attribute to FeH. This feature is similar to what we observe for KELT-20 b but our retrievals do not confidently detect FeH. Blueward of $1.1~\mu$m, there is some scatter in the measured transit depths of KELT-20 b and WASP-178 b but no discernible absorption features are present. Overall, the NIR data displays fairly weak absorption features, potentially due to the dissociation of water and/or continuum opacity from H$^{-}$ \citep{Wakeford2020}.

\section{Discussion}
\label{sec:discussion}

\subsection{Comparison to GCMs}
\label{sec:gcm}
In \S~\ref{sec: WLC and Limb}, we place upper bounds on the difference in the size (and therefore temperature-pressure profiles) of the two limbs in different wavelength bins. In order to interpret these upper bounds, we compare them to predictions from General Circulation Models (GCMs). Given the high irradiation temperature of this planet, a significant fraction of the dayside may be thermally ionized. If a magnetic field were present, these ions would resist flow across these field lines and alter the global temperature structure. We therefore run 3D numerical models with and without magnetic effects to determine if the lack of limb asymmetry found in the data could be reproduced via the inclusion of spatially varying drag. Specifically, we use the RM-GCM \citep{Rauscher2012GCM} with the updated ``picket fence'' radiative transfer scheme implemented first in \cite{Malsky2024PF} and the kinematic MHD approach, which has previously been applied to hot and ultrahot Jupiters \citep{RauscherMenou2013magdrag, Beltz2022}. Each GCM presented here spans from 100 bar to 10$^{-5}$ bar at a T31 resolution, which corresponds to roughly 3$^{\circ}$ spacing on the equator. Each model is run until it reaches a quasi-steady state at 1000 planetary orbits. We take our values for the planet radius, period, and semi-major axis from \cite{Lund2017}. There are no published measurements of KELT-20 b's mass, so we adopt a value of 2 M$_{\rm Jup}$. This is consistent with the published 3~$\sigma$ upper limit of 3.38 M$_{\rm Jup}$ \citep{Lund2017} and compatible with the value retrieved in \S~\ref{sec:petit_retrievals}.

We present two models here, a drag-free/0 Gauss model and a 5 Gauss active drag model, which differ in their treatment of drag in the atmosphere. The RM-GCM is unique in its ability to apply kinematic MHD or ``active'' drag. This approach, described first in \cite{Perna2010magdrag} and implemented for hot Jupiters in \cite{RauscherMenou2013magdrag} and ultra hot Jupiters in \cite{Beltz2022}, is a spatially varying Rayleigh drag. From a physical standpoint, this drag approximates the Lorentz force felt by thermally ionized atmospheric species as they are blown across magnetic field lines from the planet's strong winds. We assume that the planet's magnetic field is a dipole aligned with the axis of rotation. This results in drag being applied solely in the east west direction through the following timescale:
\begin{equation} \label{tdrag}
    \tau_{mag}(B,\rho,T, \phi) = \frac{4 \pi \rho \ \eta (\rho, T)}{B^{2} |sin(\phi) | }
\end{equation}
where $B$ is the chosen global magnetic field strength (in this case, 5 Gauss), $\phi$ is the latitude, $T$ is the local temperature, $\rho$ is the density for each model grid point and $\eta$ represents magnetic resistivity:
\begin{equation} \label{resistivity}
    \eta = 230 \sqrt{T} / x_{e} \textnormal{ cm$^{2}$ s$^{-1}$},
\end{equation}
where the ionization fraction $x_{e}$ can be calculated using the Saha equation. For numerical stability, a minimum timescale of 0.0025 of the planet's rotational period is imposed. 

\begin{figure}
    \includegraphics[width=\linewidth]{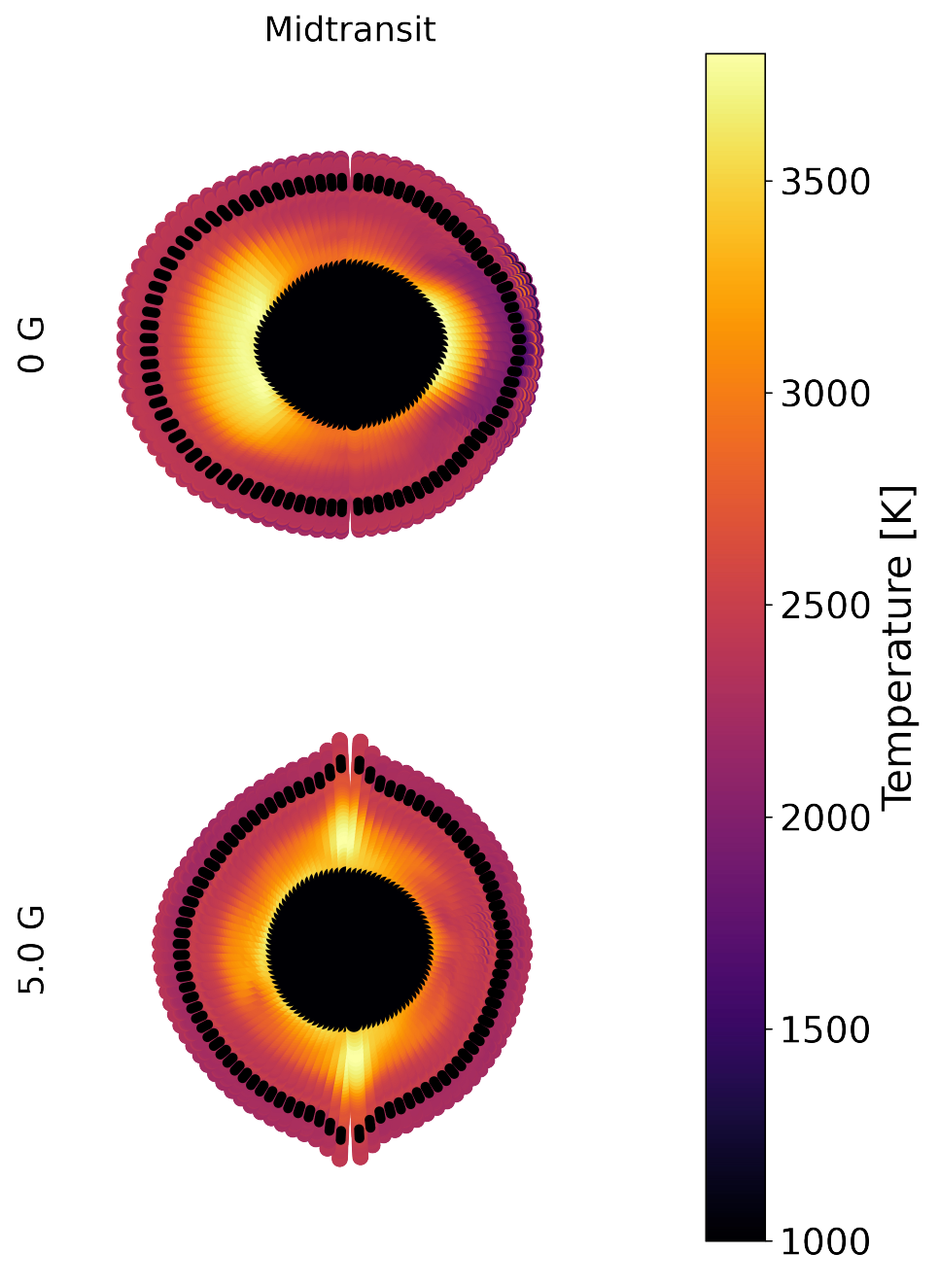}
    \caption{East (left hemisphere) and west (right hemisphere) limbs from the two GCMs presented in this work, plotted in altitude space. The black hash marks correspond to a pressure of roughly $10^{-4}$ bars. The 5G model has smaller limb asymmetries than the 0G model and is favored by the data.}
    \label{fig:GCMlimbs}
    \end{figure}

The inclusion of active magnetic drag impacts the global temperature structure of the modeled atmosphere. Most significantly, the inclusion of active drag reduces the eastward hotspot offset and can increase the day-night temperature contrast. The temperature structure of the limbs is affected, as shown in the left panel of Figure~\ref{fig:tpprofiles}, where time-averaged profiles for the east and west limb, substellar, and antistellar point are shown. As expected, the active drag model (shown with dashed lines) has a hotter dayside and a cooler nightside than the drag-free model. The eastern limb is cooler in the 5G model than the 0G model due to the active drag reducing equatorial wind speeds and decreasing heat redistribution efficiency. For the western limb, the active model is hotter in the upper atmosphere probed by the transmission spectra. The difference in the limb temperatures is much lower for the 5G model than the 0G model, which leads to different predictions for the level of limb asymmetry expected for the two models. This is visualized in Figure~\ref{fig:GCMlimbs}, where we show the temperature structure for the east and west limbs of both models. From this figure, it is apparent that the limbs of the active drag model are more symmetric than the drag-free model. 

In the right panel of Figure~\ref{fig:tpprofiles}, we compare the GCM temperature-profiles for the east and west limbs with the retrieved profiles from \texttt{PETRA}, the 1D radiative-convective equilibrium (RCE) model from \texttt{PHOENIX} assuming full heat redistribution, and condensation curves for Fe and MgSiO$_3$ from \cite{Visscher2010}. The GCM limb temperatures are much cooler in the upper atmosphere than those obtained from the retrievals and the 1D RCE model (see also \S~\ref{sec:disc:hires}). The photospheres for both the \texttt{PHOENIX} 1D RCE and the GCM profiles are at $\sim 10$ mbar. We expect the hotter profiles from the retrievals and 1D RCE to be more representative of the true temperature-pressure profile of the planet in the upper atmosphere (see also \citealt{Borsa2022, Yan2022, Singh2024} for constraints on dayside temperatures and pressures). This difference in the profiles is likely driven by the much lower NUV-optical opacity in the GCMs, which drives the much weaker inversions seen in the GCM temperature-pressure profiles. In addition, the picket fence radiative transfer scheme used for the GCMs relies on coefficients calculated for FGK stars rather than A stars such as KELT-20. 

With these caveats in mind, we use the limb profiles from the GCMs to generate spectra with \texttt{PHOENIX} and compare the predicted transit depths for the two limbs. As shown in the right panel of Figure~\ref{fig:tpprofiles}, the east limbs for the 0G and 5G models are just warm enough to avoid the condensation of silicates (and Fe for the 0G model) while the western limbs are cool enough for these species to condense out. The 0G model yields a difference in transit depth of 801 ppm and 992 ppm for the $0.2-0.8~\mu$m and $0.2-0.4~\mu$m bandpasses and are incompatible with the measured limb asymmetries (\S~\ref{sec: WLC and Limb} and Figure~\ref{fig:limb_asymmetry}) in these bandpasses at the $\sim 2.6~\sigma$ and $1.4~\sigma$, respectively. The corresponding values for the 5G model in the $0.2-0.8~\mu$m and $0.2-0.4~\mu$m bandpasses are 251 ppm ($0.5~\sigma$) and 316 ppm ($0.5~\sigma$), respectively. The data show a preference for the 5G model. In the future, this preference could be verified by measuring the location of the hot spot on the planet's day side \citep[e.g.,][]{Coulombe2023}; if magnetic drag dominates the atmospheric circulation, we would expect to see that the hot spot is located near or even westward of the substellar point \citep{Rogers2017}, rather than eastward as predicted by the 0G model.

We note that the GCM limb temperatures are likely underestimated and bringing their predicted temperatures closer to the expected/retrieved values could affect the predicted limb asymmetry in a non-trivial manner. For example, if both limbs are hot enough to prevent the condensation of the observed refractories, the predicted limb asymmetry may decrease even if the absolute difference in the limb temperatures remains the same. Future work should explore suites of GCM models with sophisticated radiative transfer and treatment of magnetic effects for UHJs.

\subsection{Comparison with high-resolution spectroscopic measurements}\label{sec:disc:hires}
Our low spectral resolution WFC3 observations indicate the presence of Fe II and/or SiO, and water in KELT-20 b's atmosphere. Under the assumption of chemical equilibrium, we find that KELT-20 b's atmosphere has a sub-solar [Z/H]. \cite{Gandhi2023} perform retrievals on six ultra hot Jupiters, including KELT-20 b, and quantify the Fe I/H abundance of these planets. For KELT-20 b, they find Fe I/H = $-0.36^{+0.35}_{-0.27} \times$ solar, which is compatible with our measured [Z/H] in the chemical equilibrium retrievals from both \texttt{PETRA} ([Z/H] = $-0.75 \pm 0.13$) and \texttt{petitRADTRANS} ([Z/H] = $-1.25^{+0.18}_{-0.15}$). However, given that they only measure the abundance of Fe I and since a significant amount of iron can be ionized at high temperatures and low pressures (Figure~\ref{fig:h2o_abund_compare}), we evaluate what the combination of low and high resolution observations can tell us about the [Z/H] of KELT-20 b. 

We also investigate if the measured iron abundance reflects the composition of the planetary envelope or if it is affected by atmospheric processes that cap its abundance in the part of the atmosphere visible to our instruments. Close-in planets such as ultra hot Jupiters are tidally locked to their host stars and therefore have permanent night sides that are significantly cooler than the planet's equilibrium temperature. If refractory species such as iron condense out into cloud particles on the night side, settle deeper into the atmosphere, and vertical mixing is weak enough to not re-mix the atmosphere on timescales shorter than advection timescale, then our measured abundance of iron would be lower than the atmosphere's true iron abundance. Phase-curve observations of hot Jupiters also suggest that photospheric night-side temperatures of these planets are nearly independent of their dayside temperatures \citep{Beatty2019, Keating2020} and refractory condensation may be acting as a thermostat on the night-side \citep{Gao2021}. Similar temperatures on the night-side could therefore provide a compelling argument for the nearly solar and narrow range of Fe I abundances measured on ultra hot Jupiters.

\begin{figure}
    \centering
    \includegraphics[width=\linewidth]{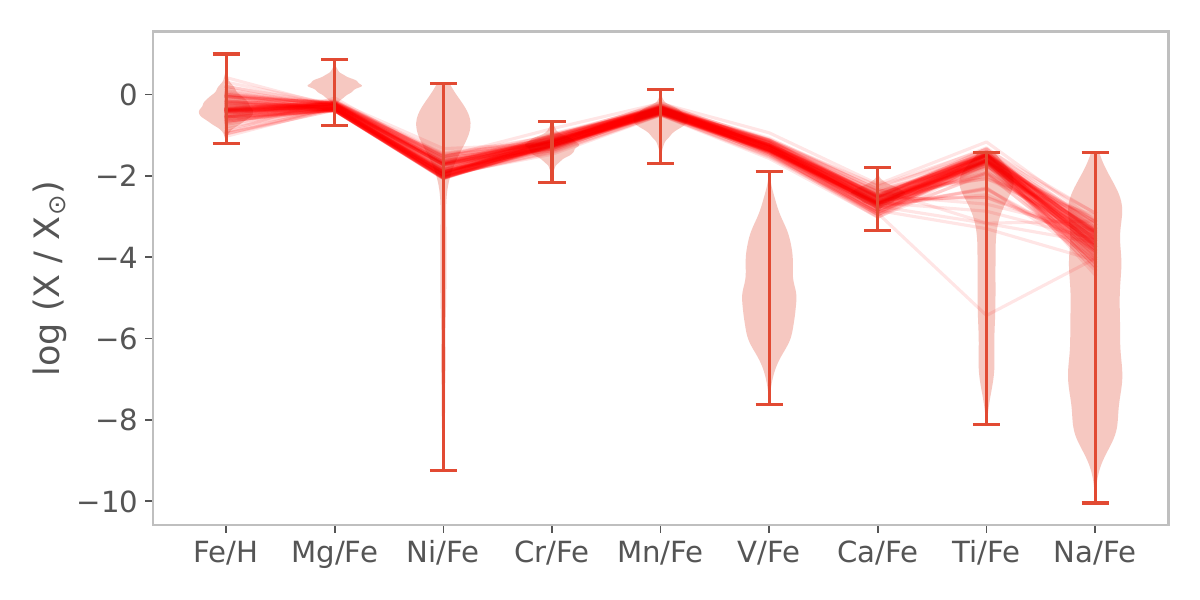}
    \caption{The measured elemental abundance ratios (violin plots) from \cite{Gandhi2023} and the fitted values from our models (red lines). All ratios are normalized to the solar value. Our model is able to account for all the observed elemental ratios with the exception of vanadium, which appears to be depleted beyond our model predictions.}
    \label{fig:high_res_fit_samples}
\end{figure}

Given that the temperature and pressure at which a condensed species is quenched determines its abundance in the photosphere, simply attempting to fit a single measurement of Fe I abundance does not yield a unique solution for the true Fe abundance and the quench temperature and pressure. One can always find a combination of these parameters that would fit the singular measurement. To truly evaluate the explanatory power of this hypothesis, it must account for the abundance of other species affected by night-side condensation as well. We therefore use the absolute Fe I abundance and the relative abundance of 7 other refractory elements from \cite{Gandhi2023} to determine if night-side condensation is throttling refractory abundance in the photosphere. We note that the abundances reported by \cite{Gandhi2023} are only for the atomic abundance of a given element (except Titanium, the reported abundance is the sum of Ti, TiO, and TiH); they do not include other states that these elements may be present in (e.g., ionized form). 

\begin{figure}
    \centering
    \includegraphics[width=\linewidth]{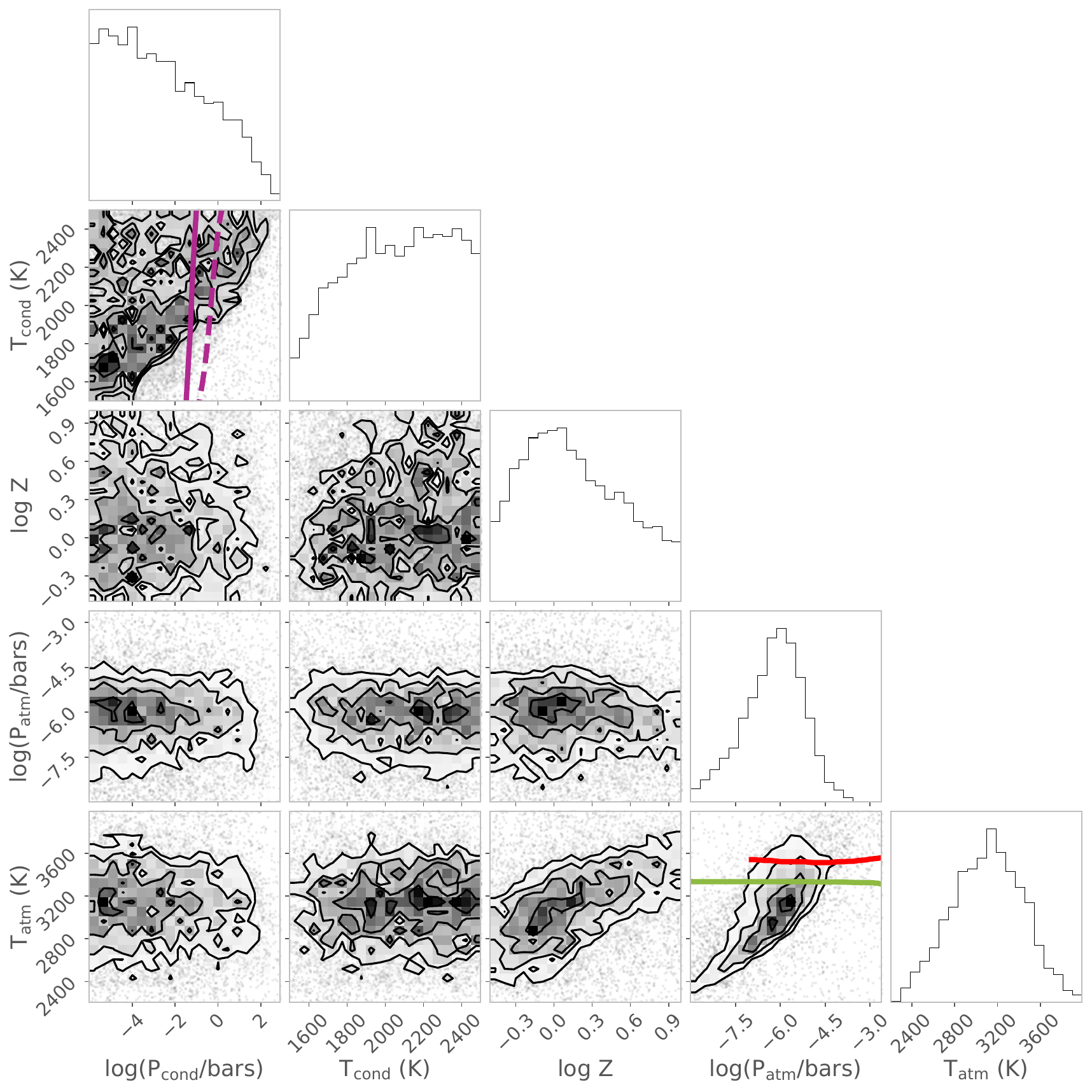}
    \caption{The posteriors for the parameters fitted to the elemental abundance ratios from \cite{Gandhi2023}. The red line shows the median temperature-pressure profile from \cite{Gandhi2023}. The other colored lines correspond to temperature-pressure profiles from Figure~\ref{fig:tpprofiles}, with magenta indicating the nightside from GCMs (solid: 0G, dashed: 5G) and green for the median profile from the chemical equilibrium retrieval with \texttt{PETRA}.
    }
    \label{fig:high_res_fit_corner}
\end{figure}

To fit these measurements, we create a grid of chemical equilibrium models using \texttt{GGChem} \citep{Woitke2018} using a two-step process. First, we calculate equilibrium abundances for a range of temperature ($T_{\rm cond} \in [1500-2500]$ K in 100 K step) and pressure ($P_{\rm cond} \in [10^{-6} - 10^{3}]$ bars, 1 dex step) that correspond to the night-side while varying the metallicity (Z/H$\in [-0.5, 1.0]$ with 0.25 step, all refractory species scaled by a single number) and the C/O ($0.2 - 1.2$ with 0.2 step) of the atmosphere. We then take the abundance of each species as the starting point for equilibrium chemistry calculations on the planet terminator, for which we vary the photospheric temperature $T_{\rm atm}$ ($1200 - 4000$ K) and pressure $P_{\rm atm}$ ($10^{-9} - 10^3$ bars). This grid is interpolated and incorporated into a nested sampling framework and compared with the posteriors of \cite{Gandhi2023} to determine the parameters that best fit the observations. We use kernel density estimation (KDE) for the posteriors from \cite{Gandhi2023} to calculate the likelihood of a given abundance from our model grid. We assume uniform priors for each parameter. 

\begin{figure*}
    \centering
    \includegraphics[width=\linewidth]{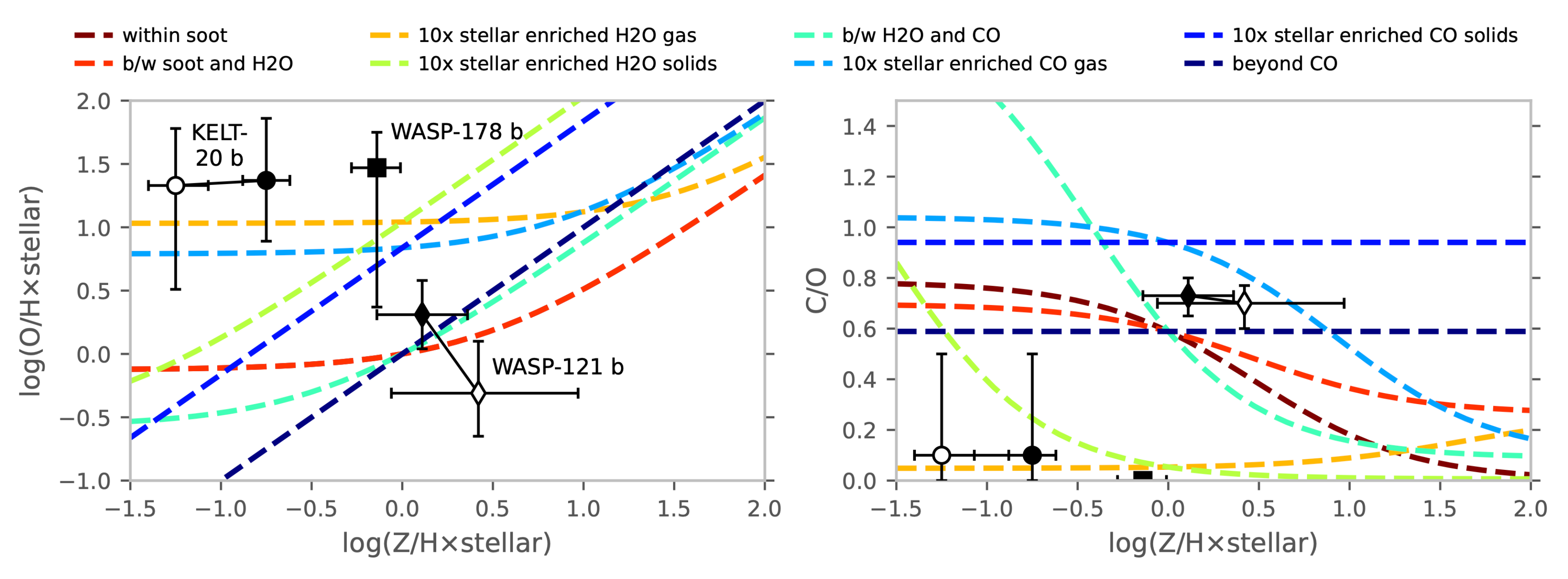}
    \caption{Compositional constraints for KELT-20 b from this work (solid circle and empty circle correspond to chemical equilibrium retrievals with \texttt{PETRA} and \texttt{petitRADTRANS}, respectively. The C/O = $0.1^{+0.4}_{-0.1}$ is from \citealt{Finnerty2025}), WASP-178 b (solid square) from \cite{Lothringer2025}, and WASP-121 b from \citep[][solid diamond]{Pelletier2025} and \citep[][empty diamond]{Smith2024}. The composition tracks corresponding to accretion in different regions of the disk are based on the models in \cite{Chachan2023}. For the sake of comparison, we assume solar composition for the host stars but note that these tracks depend on the stellar elemental ratios and would be different for WASP-121.}
    \label{fig:formation-composition}
\end{figure*}

Figure~\ref{fig:high_res_fit_samples} shows the measured and fitted abundances and Figure~\ref{fig:high_res_fit_corner} shows the posteriors for our fitted parameters. There are several interesting inferences that we can draw from this analysis. Our simple model is able to reproduce the observed refractory elemental ratios in KELT-20 b, with the exception of vanadium, which appears to be depleted beyond the model prediction. Importantly, this is accomplished not by quenching from night-side condensation but simply by expectations from thermo-chemical equilibrium. We verify this by creating another grid that does not incorporate night-side condensation and fitting for only log~$Z$, $T_{\rm atm}$, and $P_{\rm atm}$, which yields a Bayes factor of $\sim 1$. The $P_{\rm cond}$ and $T_{\rm cond}$ posteriors in Figure~\ref{fig:high_res_fit_corner} are likely driven by the precise measurement of Ca abundance such that the most refractory Ca-containing species are not condensed out \citep{Lodders2002}, with the possible exception of titanium (\citealt{Woitke2018}; see also \citealt{Kasper2023, Johnson2023}). The nightside temperature-pressure profiles from our GCMs are also shown in the $T_{\rm cond}-P_{\rm cond}$ panel. For pressures $\gtrsim 10^{-2}-10^{-1}$ bar, they are consistent with the obtained posteriors. Interestingly, the GCM profiles are colder than the $T_{\rm cond}-P_{\rm cond}$ posteriors at lower pressures, which would lead to condensation of some species. However, these condensed species would re-evaporate as they sink to pressures $\gtrsim 10^{-1}$ bar and remain in gas-phase at the limbs, as observed.

The observed elemental ratios also admit a fairly wide range of values for the atmospheric metallicity log~$Z$, which scales the `true' underlying abundance of the refractory species in the atmosphere. This broad posterior for log~$Z$ is a result of the uncertainty in the temperature and pressure of the observed region: Fe I/H remains the same with increasing log~$Z$ at a higher photospheric temperature and pressure ($T_{\rm atm}$ and $P_{\rm atm}$, Figure~\ref{fig:high_res_fit_corner}). The fitted temperature profile in \cite{Gandhi2023} is nearly isothermal with a $1 \, \sigma$ constraint of $T \sim 3500 \pm 250$ K, which is compatible with our chemical equilibrium retrieval temperature of 3350 K (both median profiles shown in Figure~\ref{fig:high_res_fit_corner}). For this $1 \sigma$ range of $T_{\rm atm}$, the posteriors in Figure~\ref{fig:high_res_fit_corner} suggest that $P_{\rm atm} \sim 10^{-7.5} - 10^{-4.5}$ bars (compatible with our \emph{Hubble} observations, Figure~\ref{fig:petra_spec_fits}) and log~$Z \in [-0.35, +1.0]$. This metallicity is higher than the [Z/H] we obtain in our chemical equilibrium retrieval with \texttt{PETRA} and \texttt{petitRADTRANS}. The two estimates can be reconciled if $T_{\rm atm}$ is slightly lower ($\sim 3000$ K) but still consistent with the estimated T-P profiles in this study and \cite{Gandhi2023} at the 2~$\sigma$ level.

\subsection{Formation-Composition link}

Our WFC3 observations indicate that KELT-20 b is sub-solar in iron but super-solar in oxygen in the chemical equilibrium retrieval framework with both \texttt{PETRA} and \texttt{petitRADTRANS}. These data imply that the planet is enriched in volatiles and depleted in refractories. We combine these measurements with the C/O = $0.1^{+0.4}_{-0.1}$ reported for KELT-20 b in \citep{Finnerty2025}. The high $T_{\rm cond}$ inferred for KELT-20 b also implies that the measured C/O reflects primordial abundances since a negligible amount of O is sequestered into refractory condensates. These measurements are compared to formation models from \cite{Chachan2023} in Figure~\ref{fig:formation-composition}. Since the elemental abundances of KELT-20 are unknown, we use solar elemental abundances from \cite{Asplund2021} for these formation models. Although the weak constraints on the planet's C/O do not enable conclusive inferences, the high O/H and low Fe/H necessitate the accretion of oxygen-rich material, for example, in the form of gas or solids enriched in water (Figure~\ref{fig:formation-composition}). Due to the large volatile-to-refractory abundance ratio, water-enriched gas is more likely compared to water-enriched solids as the source of the planet's O enrichment. The enrichment of disk gas necessary to satisfy these compositional constraints is $\sim 10 \times$ solar. Grain growth, drift, and evaporation models indicate enrichment of $\lesssim 10 \times$ solar is plausible \citep[e.g.,][]{Oberg2016, Booth2017} and observations with JWST confirm that inner disks can be enriched in water \citep{Banzatti2023}. Since [O/H] is driving this required enrichment and its measurement is sensitive to the degree of dissociation of water in the planet's atmosphere, the required enrichment would be smaller if the terminator of the planet is a little cooler than the retrieved temperature profile. This would likely simultaneously reduce [O/H] and increase [Fe/H] (to compensate for the smaller scale height). A slightly cooler temperature-pressure profile would also bring the high-resolution elemental ratios and our low-resolution constraints in better agreement (see \S~\ref{sec:disc:hires}). 

Nonetheless, these composition constraints do unambiguously support the preferential accretion of O over refractories, which suggests the accretion of volatile enriched gas. We compare KELT-20 b's composition with two other well-studied ultra hot Jupiters - WASP-178 b and WASP-121 b in Figure~\ref{fig:formation-composition}. KELT-20 b's composition is qualitatively similar to that of WASP-178 b \citep{Lothringer2025}, which has a sub-solar refractory abundance, super-solar O abundance, and an extremely low C/O, implying the accretion of O-enriched material. In contrast, WASP-121 b is characterized by a high C/O \citep{Smith2024, Pelletier2025}, especially compared to its host star C/O$_\star = 0.23 \pm 0.05$ \citep{Polanski2022}, which suggests a much more C-rich origin of its building blocks. However, making any additional inference is currently impeded by the uncertainty in the volatile-to-refractory ratio of the planet -- \cite{Smith2024} find a volatile-poor atmosphere while \cite{Pelletier2025} find a volatile-rich atmosphere. Imminent JWST results for this planet will elucidate the nature of WASP-121 b's building blocks. Given that WASP-121 b and WASP-178 b are on polar orbits around their host stars and KELT-20 b is on an aligned orbit, the similarity in the composition of KELT-20 b and WASP-178 b and the large difference in the C/O of WASP-121 b compared to these two planets imply that the relationship between dynamical configuration and composition is not `one-to-one' and likely difficult to disentangle. A larger sample size of planets is therefore necessary to empirically understand this mapping between dynamical configuration and composition.

\section{Conclusions}
\label{sec:conclusions}
KELT-20 b is a unique ultra hot Jupiter with an orbit that is well-aligned with its hot host star's spin axis. In this study, we use the \emph{Hubble Space Telescope} to build a panchromatic transmission spectrum of KELT-20 b from $0.2 - 1.7 \, \mu$m. The WFC3 UVIS instrument rewards us with access to the NUV wavelength range that is critical for detecting refractories in a low-resolution spectrum. We extract the planet's spectrum from both the +1 and -1 orders and develop techniques to correct for the contamination from higher order curved spectral traces. The transmission spectrum displays an enormous rise in the planet's apparent size in the NUV bandpass and muted absorption features in the NIR. The NUV feature is best fit by a combination of Fe II and SiO. The weak NIR absorption features suggest the presence of water. 

Chemical equilibrium retrievals with two different tools (\texttt{PETRA} and \texttt{petitRADTRANS}) suggest sub-solar refractory ([Z/H] = $-0.75^{+0.13}_{-0.13}$ and $-1.25^{+0.18}_{-0.15}$) and super-solar oxygen ([O/H] = $1.37^{+0.49}_{-0.48}$ and $1.33^{+0.45}_{-0.82}$) abundances. The latter is driven primarily by extensive thermal dissociation of water at the inferred pressures and temperatures of the terminator - even though the absorption feature's strength and the retrieved abundance in the free retrieval indicate local actual water abundance that is strongly sub-solar. We leverage the large NUV feature to detect hints of asymmetry in the two planetary limbs but do not find any significant differences within the precision of the light curve. Using GCM models with and without magnetic fields, we show that differences in limb temperatures are smaller if the planet has a magnetic field - a result that could potentially explain the observed lack of limb asymmetry.

We compare the overall refractory abundance constraints obtained from chemical equilibrium retrievals with the refractory elemental abundances presented in \cite{Gandhi2023}. By building a two-step grid of chemical equilibrium models, we show the elemental abundance ratios of refractory species are not set by nightside condensation and quenching but are likely a result of thermo-chemical equilibrium at high temperature and low pressure. The high-resolution observations yield a wide constraint on the overall refractory metallicity that is degenerate with the photospheric temperature and pressure. This constraint is compatible with our chemical equilibrium refractory metallicity for photospheric temperatures of $\sim 3000$ K, which is lower than the median temperature values in our retrievals and \cite{Gandhi2023} but within the $2 \sigma$ range. The discovery of refractory absorption in NUV in KELT-20 b ($T_{\rm eq} \sim 2250$ K) further limits the temperature range over which iron and silicate condensation takes place in hot Jupiter atmospheres (HAT-P-41 b with a $T_{\rm eq} \sim 1950$ K has no NUV excess, \citealt{Wakeford2020}).

The chemical equilibrium constraints suggest that KELT-20 b accreted volatile rich material and its solid-to-gas accretion rate was low during envelope accretion. These measurements combined with the C/O of $0.1^{+0.4}_{-0.1}$ from \cite{Finnerty2025} suggest that the accreted material was O-rich, likely in form of water-enriched gas. A tighter C/O constraint would be extremely helpful in further constraining the nature of the planet's building blocks. KELT-20 b's composition bears a qualitative resemblance to the composition of WASP-178 b - an ultra hot Jupiter on a polar orbit \citep{Lothringer2025}. Additionally, the low C/O of KELT-20 b and WASP-178 b contrast with the super-stellar C/O reported for WASP-121 b, which is also on a polar orbit \citep{Smith2024, Pelletier2025}. These observations suggest that different dynamical configurations of these planets do not map to differences in their envelope compositions in a simple dichotomous way. Comparisons of refractory and volatile composition of a larger sample of ultra hot Jupiters would facilitate the discovery of any empirical clustering or trends in their compositions.

\begin{acknowledgments}
We are grateful to the referee for their helpful suggestions. We thank Sid Gandhi for sharing the posteriors from his fit to the high resolution data of KELT-20 b and Luca Fossati for sharing the LTE and NLTE models from their work. This work is based on observations from the Hubble Space Telescope, operated by AURA, Inc. on behalf of NASA/ESA. Support for this work was provided by NASA through Space Telescope Science Institute grant GO-17082. The authors gratefully acknowledge the computing time granted by the Resource Allocation Board and provided on the supercomputer Emmy at NHR-Nord@G{\"o}ttingen as part of the NHR infrastructure. The calculations for this research were conducted with computing resources under the project hhp00051. The \emph{Hubble} data presented in this article were obtained from the Mikulski Archive for Space Telescopes (MAST) at the Space Telescope Science Institute. The specific observations analyzed can be accessed via \dataset[doi: 10.17909/85yb-4k35]{http://dx.doi.org/10.17909/85yb-4k35}.
\end{acknowledgments}

\vspace{5mm}

\facilities{HST (WFC3)}

\software{astropy \citep{2013A&A...558A..33A, 2018AJ....156..123A}, batman \citep{Kreidberg2015}, catwoman \citep{Jones_catwoman}, LDTk \citep{Parviainen2015}, Matplotlib \citep{Hunter2007}, NumPy \citep{oliphant2006guide, van2011numpy}, petitRADTRANS \citep{molliere_2019}, PETRA \citep{Lothringer2020_petra}  
}

\appendix

\section{Retrieval priors and results}

The free parameters, priors, and measurements from our retrievals are listed in Tables~\ref{tab:retrieval_comparison} and \ref{tab:prt_retrieval_comparison}.

\begin{deluxetable}{lcccc}
\tablecaption{Comparison of Retrieval Constraints from PETRA \label{tab:retrieval_comparison}}
\tablehead{
\colhead{\uline{Parameter}} & \multicolumn{2}{c}{\uline{Chemical Equilibrium}} & \multicolumn{2}{c}{\uline{Free Chemistry}} \\
\colhead{} & \colhead{Prior} & \colhead{Measurement} & \colhead{Prior} & \colhead{Measurement}
}
\startdata
$\beta$  & $\mathcal{U}$(0.25, 1.75)  & $1.26^{+0.21}_{-0.44}$ & $\mathcal{U}$(0.25, 1.75) & $1.34^{+0.09}_{-0.10}$ \\
$\log_{10}\gamma_1$ & $\mathcal{U}$(-2, 2) & $0.43^{+0.88}_{-0.48}$ & $\mathcal{U}$(-2, 2)  & $1.79^{+0.13}_{-0.19}$ \\
$\kappa_{\rm IR}$ & $\mathcal{U}$(-4, 1) & $-3.55^{+0.98}_{-0.72}$ & $\mathcal{U}$(-4, 1) & $-0.72^{+32}_{-0.27}$ \\
$R_p$ ($R_J$)  & $\mathcal{U}$(0.14, 3.5) & $1.648^{+0.037}_{-0.034}$ & $\mathcal{U}$(0.14, 3.5) & $1.708^{+0.007}_{-0.008}$ \\
Data Offset (ppm)  & $\mathcal{U}$($10^3$, $10^3$) & $-143^{+30}_{-31}$ & $\mathcal{U}$($10^4$, $10^4$) & $11^{+11}_{-11}$ \\
$\mathrm{[Fe/H]}$  & $\mathcal{U}$(-1.0, 2.5) & $-0.75^{+0.13}_{-0.13}$ & $\mathcal{U}$(-2, 2) & $-1.02^{+0.67}_{-0.56}$ \\
$\mathrm{[O/H]}$  & $\mathcal{U}$(-2.0, 1.75) & $1.37^{+0.49}_{-0.48}$ & - & - \\
Fe II & - & - & $\mathcal{U}$(-12, -1) & $-2.47^{+0.35}_{-0.53}$ \\
TiO  & - & - & $\mathcal{U}$(-12, -1) & $-9.52^{+0.46}_{-0.59}$ \\
e$^-$  & - & -  & $\mathcal{U}$(-12, -1) & $-4.93^{+0.51}_{-0.37}$ \\
Fe  & - & - & $\mathcal{U}$(-12, -1) & $-9.36^{+0.93}_{-0.98}$ \\
SiO  & - & - & $\mathcal{U}$(-12, -1) & $-6.08^{+0.58}_{-0.62}$ \\
Mg II  & - & - & $\mathcal{U}$(-12, -1) & $-4.96^{+1.45}_{-2.41}$ \\
Mg  & - & - & $\mathcal{U}$(-12, -1) & $-5.45^{+0.66}_{-0.58}$ \\
H$_2$O  & - & -  & $\mathcal{U}$(-12, -1) & $-4.34^{+0.23}_{-0.19}$ \\
\enddata
\end{deluxetable}

\begin{deluxetable}{lcccccc}
\tablecaption{Comparison of Retrieval Constraints from \texttt{petitRADTRANS} retrievals \label{tab:prt_retrieval_comparison}}
\tablehead{
\colhead{\uline{Parameter}} & \multicolumn{2}{c}{\uline{Chemical Equilibrium}} & \multicolumn{2}{c}{\uline{Free Chemistry w/ SiO}} & \multicolumn{2}{c}{\uline{Free Chemistry w/o SiO}} \\
\colhead{} & \colhead{Prior} & \colhead{Measurement} & \colhead{Prior} & \colhead{Measurement}
 & \colhead{Prior} & \colhead{Measurement}}
\startdata
$P_{ref}$ (bars)  & fixed & 0.01 & fixed & 0.01 & fixed & 0.01 \\
$R_{*}$ ($R_{\odot}$)  & fixed & 1.6 & fixed & 1.6 & fixed & 1.6 \\
$R_p$ ($R_J$)  & $\mathcal{U}$(1.0, 3.0) & $1.827^{+0.005}_{-0.005}$ & $\mathcal{U}$(1.0, 3.0) & $1.786^{+0.007}_{-0.007}$ & $\mathcal{U}$(1.0, 3.0) & $1.794^{+0.006}_{-0.006}$ \\
$log_{10}~g$ [cgs] & $\mathcal{U}$(2.8, 4.5)  & $3.08^{+0.05}_{-0.04}$ & $\mathcal{U}$(2.8, 4.5) & $3.53^{+0.08}_{-0.08}$ & $\mathcal{U}$(2.8, 4.5) & $3.27^{+0.07}_{-0.07}$\\
$T_{eq}$ [K] & $\mathcal{U}$(1800,5500)  & $4182^{+137}_{-190}$ & $\mathcal{U}$(1800,5500) & $4620^{+600}_{-680}$ & $\mathcal{U}$(1800,5500) & $4260^{+660}_{-590}$\\
$T_{int}$ [K] & fixed  & 200 & fixed & 200 & fixed & 200 \\
$\log_{10}\gamma_1$ & $\mathcal{U}$(-3, 2) & $1.70^{+0.21}_{-0.72}$ & $\mathcal{U}$(-3, 2)  & $1.72^{+0.26}_{-0.20}$ & $\mathcal{U}$(-3, 2)  & $1.83^{+0.1}_{-0.15}$ \\
$log_{10}~\kappa_{\rm IR}$ & $\mathcal{U}$(-6, 4) & $-0.69^{+0.42}_{-0.22}$ & $\mathcal{U}$(-6, 4) & $-0.34^{+0.40}_{-0.44}$ & $\mathcal{U}$(-6, 4) & $-2.92^{+0.23}_{-0.20}$ \\
$\mathrm{[Fe/H]}$  & $\mathcal{U}$(-2.0, 2.5) & $-1.25^{+0.18}_{-0.15}$ & - & - & - & -\\
$\mathrm{[O/H]}$  & $\mathcal{U}$(-2.0, 3) & $1.33^{+0.45}_{-0.82}$ & - & - & - & - \\
$log_{10}$ e$^-$ [VMR] & - & -  & $\mathcal{U}$(-12, 0) & $-3.15^{+1.03}_{-2.03}$ & $\mathcal{U}$(-12, 0) & $-4.90^{+1.9}_{-2.0}$ \\
$log_{10}$ H [VMR] & - & -  & $\mathcal{U}$(-12, 0) & $-3.23^{+1.00}_{-0.96}$ & $\mathcal{U}$(-12, 0) & $-4.85^{+1.9}_{-2.0}$ \\
$log_{10}$ Fe II [VMR] & - & - & $\mathcal{U}$(-14, 0) & $-8.49^{+2.78}_{-2.48}$ & $\mathcal{U}$(-14, 0) & $-2.74^{+0.25}_{-0.24}$ \\
$log_{10}$ Fe [VMR] & - & - & $\mathcal{U}$(-14, 0) & $-8.67^{+1.95}_{-2.10}$ & $\mathcal{U}$(-14, 0) & $-8.12^{+0.72}_{-1.03}$ \\
$log_{10}$ SiO [VMR] & - & - & $\mathcal{U}$(-14, 0) & $-3.74^{+0.17}_{-0.23}$ & - & - \\
$log_{10}$ TiO [VMR] & - & - & $\mathcal{U}$(-14, 0) & $-11.01^{+1.38}_{-1.31}$ & $\mathcal{U}$(-14, 0) & $-10.20^{+0.18}_{-0.28}$ \\
$log_{10}$ Mg II [VMR] & - & - & $\mathcal{U}$(-14, 0) & $-7.55^{+3.37}_{-3.27}$ & $\mathcal{U}$(-14, 0) & $-6.8\pm1.8$\\
$log_{10}$ Mg [VMR] & - &  & $\mathcal{U}$(-14, 0) & $-6.55^{+3.02}_{-3.02}$ & $\mathcal{U}$(-14, 0) & $-6.7\pm2.0$ \\
$log_{10}$ H$_2$O [VMR] & - & - & $\mathcal{U}$(-14, 0) & $-4.07^{+1.15}_{-4.04}$  & $\mathcal{U}$(-14, 0) & $-5.62^{+0.15}_{-0.17}$ \\
$log_{10}$ K [VMR] & - & - & $\mathcal{U}$(-14, 0) & $-3.68^{+0.81}_{-2.67}$ & $\mathcal{U}$(-14, 0) & $-7.47^{+0.51}_{-0.81}$ \\
\enddata
\end{deluxetable}

\bibliography{sample631}{}
\bibliographystyle{aasjournal}

\end{document}